%% file: main.tex
\documentclass[runningheads]{llncs}

\usepackage{eccv}

\usepackage{eccvabbrv}

\usepackage{graphicx}
\usepackage{booktabs}

\usepackage[accsupp]{axessibility}  
\input{preamble}

\usepackage{hyperref}

\usepackage{orcidlink}

\begin{document}

\title{StructureGS: Structure-aware Gaussian Splatting for Articulated Object Reconstruction} 

\titlerunning{StructureGS}

\author{Gahye Lee\orcidlink{0009-0006-3573-4954} \and
Gyoonseo Kim\orcidlink{0009-0002-1842-2757} \and 
Wonjong Jang\orcidlink{0000-0002-1442-9399}\and 
Jooeun Son\orcidlink{0009-0003-6916-9912}\and  \\
Seungyong Lee\orcidlink{0000-0002-8159-4271}} 

\authorrunning{G. Lee et al.}

\institute{POSTECH, South Korea \\ 
\email{\{gahye0509, nevermore, wonjong, jeson, leesy\}@postech.ac.kr}\\
}

\maketitle

\input{sec/0_abstract}    
\input{sec/1_intro}

\input{sec/2_related_work}

\input{sec/3_BoundedGS}
\input{sec/4_Optimization}
\input{sec/5_Experiments}

\input{sec/6_Conclusion}

\section*{Acknowledgements}
We thank anonymous reviewers for their valuable feedback. This work was supported by NRF grants (RS-2026-25485382, RS-2025-02216257), IITP grant (RS-2022-II220290), KOCCA grant (RS-2026-25508598) funded by the Korean government (MSIT \& MCST).


%
%
\bibliographystyle{splncs04}
\bibliography{main}

\newpage
\section{Supplementary Material}
\input{sec/X_suppl}

\end{document}

%% file: preamble.tex
\usepackage{tikz-cd}

\newcommand{\Fig}[1] {Fig.~\ref{fig:#1}}

\newcommand{\Tbl}[1]  {Table~\ref{tbl:#1}}

\usepackage{graphicx}
\usepackage{fmtcount}
\usepackage{epsfig} 
\usepackage{amsmath}

\usepackage{amssymb}  
\usepackage{booktabs} 
\usepackage{subcaption} 
\usepackage{lipsum}
\usepackage{arydshln} 
\usepackage{comment}
\usepackage{algorithm} 
\usepackage{algpseudocode}  
\usepackage{bm}
\usepackage[cjk]{kotex}
\usepackage[table]{xcolor}
\definecolor{ourscolor}{RGB}{240,240,240} 
\usepackage{xcolor}
\usepackage{color, soul}
\usepackage{multirow}
\usepackage[table]{xcolor}
\usepackage{cuted}   
\usepackage{capt-of} 
\usepackage{wrapfig}
\usepackage{subcaption}
\usepackage{makecell}
\usepackage{pifont}

\usepackage{tikz}

%% file: sec/0_abstract.tex
\begin{abstract}
Reconstructing articulated objects with multiple movable parts is essential for understanding object structure and enabling physical interaction. However, this reconstruction task poses significant challenges due to the entanglement of geometry, appearance, and motion parameters during optimization. Existing methods rely primarily on photometric supervision, which commonly fails to disentangle these interdependent components, resulting in poor part decomposition with blurred boundaries and geometric artifacts. To address this limitation, we introduce {\em StructureGS}, a reconstruction framework for articulated objects that integrates structure-aware guidance into 3D Gaussian Splatting.
Our approach leverages oriented bounding boxes of object parts to enforce two key structural properties: 
{\em spatial coherence}, which constrains each part’s geometry to remain compact and spatially coherent within its designated region,
and {\em structural connectivity}, which enforces physically plausible contact relationships between adjacent parts.
These properties are realized through structure-aware losses that inject explicit structural constraints into the optimization process.
Extensive experiments demonstrate that our method achieves state-of-the-art performance in articulated object reconstruction, producing high-quality results with well-defined part geometries.
\end{abstract}

%% file: sec/1_intro.tex
\section{Introduction}
\label{sec:intro}

Articulated objects with multiple movable parts are omnipresent in human-made environments. Accurate reconstruction of these objects is crucial for embodied AI systems~\cite{deitke2022, shridhar2020alfred, szot2021habitat} to correctly understand object structure and plan physical interaction. Beyond capturing static geometry, such reconstruction must also recover the underlying kinematic structure, which defines how different parts move relative to each other with constraints on their possible configurations.

Unlike typical object reconstruction~\cite{wang2018pixel2mesh, park2019deepsdf, seitz2006comparison, chang2015shapenet, mescheder2019occupancy} that focuses solely on static geometry and appearance, reconstructing objects with a kinematic structure poses significant challenges due to the need to jointly optimize multiple interdependent components: per-part \textit{geometry}, \textit{appearance}, and \textit{motion parameters}. These components are inherently entangled, as the observed appearance depends on the underlying geometry, while accurate geometry recovery relies on understanding how parts are spatially structured and move relative to each other. This entanglement often hinders accurate estimation of each component, resulting in suboptimal reconstruction quality.


\input{figs/teaser}
Recent works on articulated object reconstruction represent geometry and appearance using neural radiance fields~\cite{liu2023paris, noguchi2022watch, jiang2022ditto, heppert2023carto, wei2022self} or Gaussian splatting~\cite{kerr2024robot,lin2024gaussian, luiten2024dynamic, guo2025articulatedgs, liu2025artgs, kim2025screwsplat}, and parameterize part motions as rigid transformations~\cite{xu2009joint,abbatematteo2019learning, li2020category, qiu2025articulate, geng2023gapartnet, liu2023self}. 
These methods rely primarily on photometric reconstruction objectives to update 3D model parameters, and may recover high-quality renderings of the {\em whole} object. 
However, without explicit structural guidance, photometric supervision alone often entangles geometry, appearance, and motion parameters during reconstruction, leading to poor {\em part} decomposition with blurred boundaries and geometric artifacts, as shown in ~\Fig{teaser}.

To address this problem, we introduce structure-aware guidance for the reconstruction framework. 
In this work, we use the term {\em structure} to refer to both the geometry of each part and the contact relationships between adjacent parts, which together characterize how an articulated object maintains a physically plausible configuration.
Our structure-aware guidance enforces two properties: 
{\em spatial coherence}, which constrains each part’s geometry to remain compact and spatially coherent within its designated region,
and {\em structural connectivity}, which enforces physically plausible contact relationship between adjacent parts. 
To effectively encode structural information, we leverage an oriented bounding box (OBB) for each articulated part. The OBB explicitly defines the oriented extent of each part, providing spatial boundary information for reasoning about its geometric coherence. Furthermore, its convexity facilitates efficient distance computation, allowing for the direct evaluation of connectivity between parts.

Based on the OBB representation, we design two complementary loss functions: \textit{part fitting loss} and \textit{part contact loss}.
The part fitting loss encourages spatial coherence by 
constraining the reconstructed geometry of a part to remain within a tight bounding box.
This drives tightly bounded part representations, reducing scattered and spatially incoherent geometry.
The part contact loss promotes structural connectivity by leveraging the Separating Axis Theorem (SAT)~\cite{ericson2004real}, which identifies shape intersection by examining projection overlaps across candidate separating axes. 
By computing inter-OBB distances using SAT, our loss encourages physically plausible contact between adjacent parts.

Building upon the structure-aware losses, we present {\em StructureGS}, a novel framework for articulated object reconstruction, where each part is represented as a set of 3D Gaussian splats~\cite{kerbl20233d}.
Importantly, in our framework, the motion of each part, shared by all Gaussians belonging to that part, is parameterized and optimized using its OBB.
Given multi-view RGB images capturing an object before and after articulation, our part fitting and part contact losses, together with a photometric loss, provide explicit structural guidance for jointly optimizing per-part geometry, appearance, and motion parameters.
This structure-aware design enables high-quality reconstruction with well-defined part boundaries and physically plausible articulated configurations, and provides robustness even under sparse observations, enhancing its practical applicability in real-world capture scenarios.

In summary, our main contributions are:
\begin{itemize}
\item We present {\em StructureGS}, a framework integrating structure-aware guidance into 3D Gaussian Splatting for high-quality reconstruction of articulated objects with clear part boundaries. 

\item We introduce structure-aware guidance via two OBB-based loss functions, \textit{part fitting loss} and \textit{part contact loss}, that enforce \textit{spatial coherence} and \textit{structural connectivity} of parts during optimization.

\item 
Our method achieves state-of-the-art performance on benchmarks with clear part decomposition and accurate motion estimation. It also provides robust reconstructions with sparse views and generalizes well to real-world objects.
\end{itemize}

%% file: figs/teaser.tex
\begin{figure}[t]
    \centering
    \includegraphics[width=\linewidth,trim={0cm 9cm 0cm 0cm}]{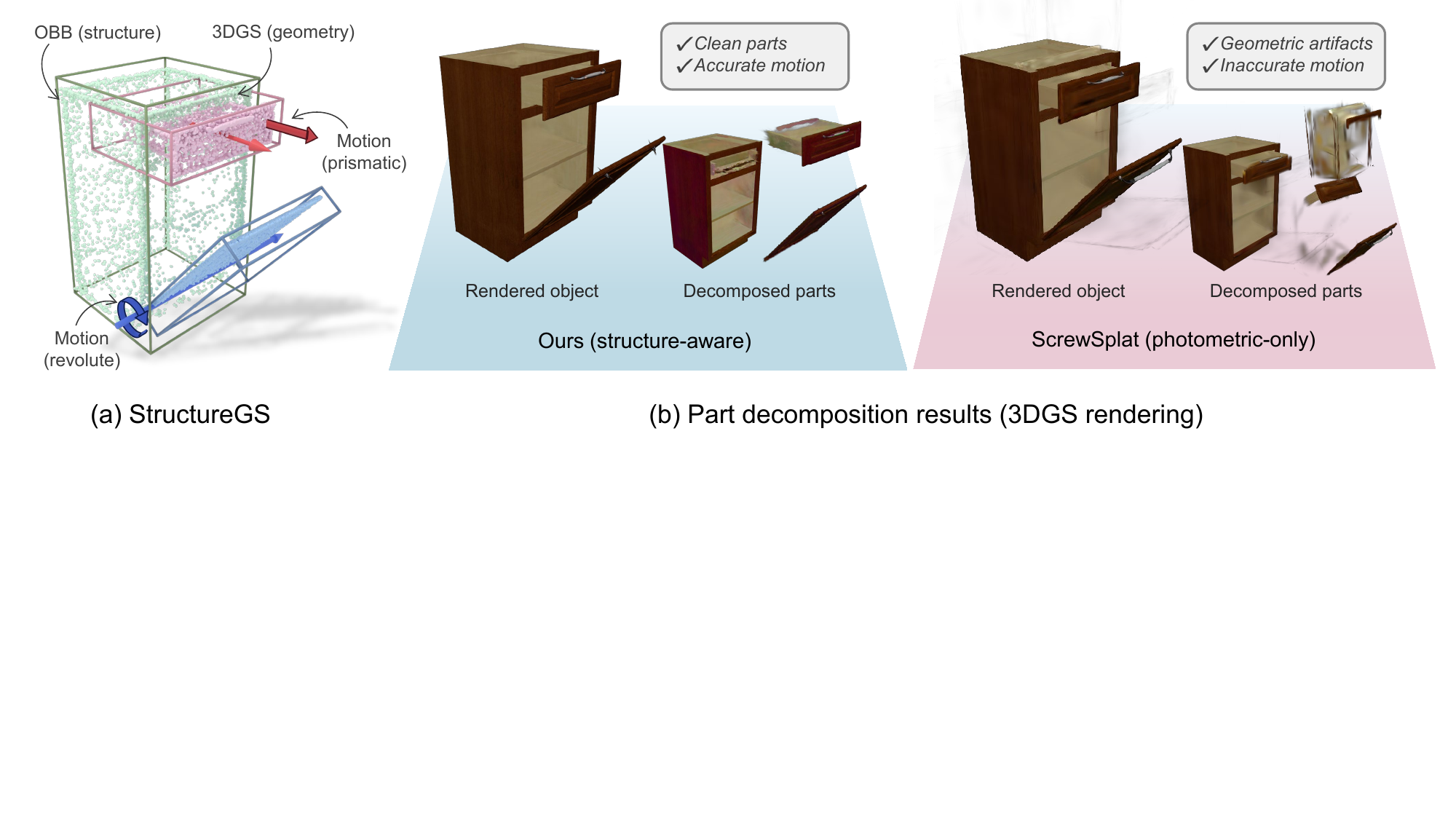}
    \caption{{\em StructureGS}. (a) Each part is represented by an explicit OBB (structure) and 3D Gaussians (geometry), with prismatic or revolute motion parameters. (b) Structure-aware optimization yields cleaner part decomposition and more accurate motion estimation than a photometric-only method (ScrewSplat~\cite{kim2025screwsplat}). }
    \label{fig:teaser}
\end{figure}

%% file: sec/2_related_work.tex
\section{Related Work}
\label{sec:relwork}

\paragraph{Articulated object reconstruction}
Articulated object reconstruction methods can 
be categorized into \emph{holistic} and \emph{part-wise} approaches. Holistic methods reconstruct the object as a single continuous representation, where part structure is implicitly captured by latent variables or deformation fields. In contrast, part-wise methods decompose the object as a set of explicit parts, each with its own geometry and motion parameters. \Tbl{method_comparison} summarizes the key characteristics of representative methods. 

Holistic deformation-based methods, such as A-SDF~\cite{mu2021sdf}, represent the object's shape and articulation within a single signed distance function by learning separate latent codes for shape and pose.
Similarly, REACTO~\cite{song2024reacto} employs a skeleton-driven deformation field to explain appearance changes across a video sequence. These approaches can produce plausible whole-object reconstructions under large deformations, but they lack explicit control over part-level geometry and often struggle to recover clean part boundaries.

Alternatively, part-wise approaches model an articulated object as a composition of multiple parts with part-level geometry and rigid motions~\cite{jiang2022ditto, liu2023paris, heppert2023carto, kim2025screwsplat}. PARIS~\cite{liu2023paris} jointly reconstructs part-level radiance fields and motions from multi-view captures. Within the realm of 3D Gaussian Splatting (3DGS), ArticulatedGS~\cite{guo2025articulatedgs} and ArtGS~\cite{liu2025artgs} discover part structures through deformation-aware clustering and skeleton-guided spectral clustering of Gaussian splats, respectively. Other recent works focus on mobility: SPLART~\cite{lin2025splart} infers part assignments from per-Gaussian mobility estimates, while ScrewSplat~\cite{kim2025screwsplat} groups Gaussians via screw-motion parameters. In this work, we adopt the part-wise paradigm, explicitly representing each movable part with a set of 3D Gaussians and an oriented bounding box to jointly recover its structure and motion.

\begin{table}[t]
  \begin{minipage}{0.6\textwidth}
    \centering
    \resizebox{\textwidth}{!}{%
    \begin{tabular}{c|c|c|cccc}
        \hline
        Method & Input & Representation
        & \rotatebox[origin=c]{90}{Multi-part}
        & \rotatebox[origin=c]{90}{Part-wise}
        & \rotatebox[origin=c]{90}{Joint type}
        & \rotatebox[origin=c]{90}{\makecell{Structure-\\awareness}} \\
        \hline
        A-SDF~\cite{mu2021sdf} & {\scriptsize PC} & \multirow{3}{*}{NeRF}
        & & & & \\
        REACTO~\cite{song2024reacto} & RGB &
        & & & & \\
        PARIS~\cite{liu2023paris} & RGB &
        & & \ding{51} & \ding{51} & \\
        \hline
        SPLART~\cite{lin2025splart} & RGB & \multirow{5}{*}{\makecell{Gaussian\\Splatting}}
        & & & & \\
        ArtGS~\cite{liu2025artgs} & RGBD &
        & \ding{51} & & \ding{51} & \\
        ScrewSplat~\cite{kim2025screwsplat} & RGB &
        & \ding{51} & \ding{51} & \ding{51} & \\
        ArticulatedGS~\cite{guo2025articulatedgs} & RGB &
        & \ding{51} & & \ding{51} & \\
        \rowcolor{yellow!30} Ours & RGB & & \ding{51} & \ding{51} & \ding{51} & \ding{51} \\
        \hline
    \end{tabular}%
    }
  \end{minipage}\hfill
  \begin{minipage}{0.36\textwidth}
    \caption{Comparison of articulated object reconstruction methods.
    \textit{Multi-part} indicates support for arbitrary $k$ parts;
    \textit{Part-wise} denotes per-part representation rather than a holistic one;
    \textit{Joint type} means no predefined joint type information is required;
    \textit{Structure-awareness} reflects the use of structural constraints beyond photometric supervision.}
    \label{tbl:method_comparison}
  \end{minipage}
\end{table}
%
\paragraph{Photometric supervision and its limitation}
Photometric supervision has been the primary objective used in previous articulated object reconstruction methods.
However, although appearance changes in articulated objects are solely induced by part motions, photometric supervision cannot distinguish whether the appearance changes have been caused by geometry change or motion. 
As a result, the optimizer often incorrectly adjusts the geometry to account for appearance changes caused by motion,
rather than accurately updating kinematic parameters,
which leads to blurred part boundaries or geometry leaking across parts.

To mitigate this ambiguity, several approaches incorporate auxiliary priors to better estimate articulation or part geometry.
Skeleton-driven methods such as REACTO~\cite{song2024reacto} use predefined joint hierarchies to guide deformation.
Clustering-based approaches including ArticulatedGS~\cite{guo2025articulatedgs}, ArtGS~\cite{liu2025artgs}, and SPLART~\cite{lin2025splart} form part groups using deformation similarity, spectral analysis, or per-Gaussian mobility.
ScrewSplat~\cite{kim2025screwsplat} constrains the space of rigid motions through a screw-motion formulation.

These methods provide valuable supplementary supervision for identifying how an articulated object moves or how its elements should be grouped.
However, they do not constrain the structure of parts, such as shapes, boundaries, and physical contacts, which is critical for robust reconstruction of articulated objects.
In this paper, we address this issue by introducing structure-aware losses that explicitly promote coherent part shapes and plausible part connectivity.

%% file: sec/3_BoundedGS.tex
\input{figs/overview}
\section{Structure-Aware Gaussian Splatting}
\label{sec:GroupGS}

Given multi-view RGB images $\{\mathbf{I}_v^t\}$ and corresponding camera parameters $\{\mathbf{C}_v\}$ of an articulated object captured in two distinct configurations, the initial state at $t=0$ and an articulated state at $t=1$, our objective is to jointly reconstruct the part-wise 3D geometry, appearance, and motion parameters that faithfully explain the observations across both states. Similar to prior articulated reconstruction works~\cite{liu2025artgs, guo2025articulatedgs, liu2023paris, lin2025splart}, we model the articulated object as $K$ rigid parts $\mathcal{P} = \{P_0, P_1, \ldots, P_{K-1}\}$, where $P_0$ is a fixed base and each movable part $P_k$ ($k \geq 1$) transforms rigidly between the two states.



\subsection{Part Representation}
\label{sec:representation}

Our part representation is designed to disentangle intrinsic geometry from extrinsic pose. Each part $P_k$ consists of Gaussian primitives in a canonical coordinate frame, coupled with transformation parameters that determine its location, orientation, and scale in the world space.

\paragraph{Canonical Gaussian primitives}
Each part $P_k$ is represented by a group $G_k$ consisting of $N_k$ Gaussian primitives defined in a normalized canonical space, along with transformation parameters $\{T_0^k, T_1^k\}$, where $T_0^k$ is an affine transformation that maps the canonical space to the world space at $t=0$, and $T_1^k$ is a rigid transformation that maps the initial state to the articulated state:
\begin{equation*}
G_k = \{(\boldsymbol{x}_j, \alpha_j, \boldsymbol{q}_j, \boldsymbol{c}_j)\}_{j=1}^{N_k} \cup \{T_0^k, T_1^k\},
\end{equation*}
where $\boldsymbol{x}_j \in \mathbb{R}^3$ is the canonical position, $\alpha_j \in [0,1]$ is the opacity, $\boldsymbol{q}_j$ parameterizes the covariance through rotation 
and scale, and $\boldsymbol{c}_j$ denotes spherical harmonic coefficients for view-dependent appearance.
To ensure that Gaussian positions remain within the normalized canonical space, we constrain them to the cube $[-1, 1]^3$ using:
\begin{equation*}
\boldsymbol{\mu}_j = \tanh(\boldsymbol{x}_j), \qquad \boldsymbol{\mu}_j \in [-1, 1]^3.
\end{equation*}
This canonical representation decouples the local geometric details of each part from its global pose and spatial extent, 
facilitating explicit modeling and optimization of part structure through bounding box parameters (\Fig{overview}(a), right).

\paragraph{Transformation to world space}
The transformation parameters $T_0^k = (\boldsymbol{s}_0^k, R_0^k, \boldsymbol{t}_0^k)$ define an affine transformation that maps the canonical Gaussians to the world coordinate system at the initial state $t=0$. For the $j$-th Gaussian in part $k$, the world position is computed as:
\begin{equation*}
\label{eq:state0}
\boldsymbol{\mu}_j^{k,0} = \boldsymbol{s}_0^k \odot (R_0^k \boldsymbol{\mu}_j) + \boldsymbol{t}_0^k,
\end{equation*}
where $\boldsymbol{s}_0^k \in \mathbb{R}_{>0}^3$ represents per-axis scaling, $R_0^k \in SO(3)$ is the rotation, and $\boldsymbol{t}_0^k \in \mathbb{R}^3$ is the translation. The element-wise product $\odot$ with the scale vector effectively parameterizes an oriented bounding box that encodes the part's spatial extent and orientation. The covariance matrix is similarly transformed as:
\begin{equation*}
\boldsymbol{\Sigma}_j^{k,0} = R_0^k \, \text{diag}(\boldsymbol{s}_0^k) \, \boldsymbol{\Sigma}_j \, \text{diag}(\boldsymbol{s}_0^k) \, (R_0^k)^\top.
\end{equation*}

\paragraph{Transformation to articulated state}
For the articulated state $t=1$, we apply a rigid transformation $T_1^k = (R_1^k, \boldsymbol{t}_1^k)$ to the initial position of part $P_k$ at $t=0$:
\begin{equation*}
    \boldsymbol{\mu}_j^{k,1} = R_1^k \boldsymbol{\mu}_j^{k,0} + \boldsymbol{t}_1^k \quad \mbox{and} \quad
    \boldsymbol{\Sigma}_j^{k,1} = R_1^k \boldsymbol{\Sigma}_j^{k,0} (R_1^k)^\top,
\end{equation*}
where $R_1^k \in SO(3)$ and $\boldsymbol{t}_1^k \in \mathbb{R}^3$ capture the part's rigid motion between states (\Fig{overview}(a), middle and left).

\subsection{Part Contactness}
\label{sec:connectivity}

A fundamental property of articulated objects is that connected parts remain in physical contact. We exploit this structural constraint by computing inter-part distances directly from the bounding box parameters equipped in our part representation.

\paragraph{Bounding box extraction}
Since each part is defined in a normalized canonical space $[ -1, 1]^3$, transformation $T_0^k = (\boldsymbol{s}_0^k, R_0^k, \boldsymbol{t}_0^k)$ naturally induces the oriented bounding box for part $P_k$ at the initial state $t=0$. The OBB is centered at $\boldsymbol{t}_0^k$ with half-extent $\boldsymbol{s}_0^k$ and orientation defined by $R_0^k$. Similarly, the OBB at the articulated state $t = 1$ can be obtained by applying $T_1^k$ to the OBB in the initial state. This direct correspondence between part parameters and bounding boxes enables efficient contact evaluation without explicit mesh extraction.

\paragraph{Distance computation via Separating Axis Theorem}
\input{figs/sat}

To compute the distance between two parts $P_i$ and $P_j$, we employ the Separating Axis Theorem~\cite{ericson2004real}, which provides an efficient method for determining the distance between two convex objects by projecting both objects onto a set of candidate axes.

For two OBBs, the complete set of candidate axes consists of 15 directions:
\begin{equation*}
\mathcal{A} = \{\boldsymbol{n}_i^{1}, \boldsymbol{n}_i^{2}, \boldsymbol{n}_i^{3}, \boldsymbol{n}_j^{1}, \boldsymbol{n}_j^{2}, \boldsymbol{n}_j^{3}\} \cup \{\boldsymbol{n}_i^{p} \times \boldsymbol{n}_j^{q}\}_{p,q=1}^3,
\end{equation*}
comprising the three face normals of each OBB (obtained from the columns of $R_0^i$ and $R_0^j$), and the nine pairwise cross products of their edge directions.
For each axis $\boldsymbol{a} \in \mathcal{A}$, we project both bounding boxes onto the normalized axis $\hat{\boldsymbol{a}} = \boldsymbol{a}/\|\boldsymbol{a}\|$, yielding one-dimensional intervals with centers $c_i$ and $c_j$ and extents $r_i$ and $r_j$. The signed distance along this axis is:
\begin{equation*}
d_{\boldsymbol{a}} = |c_i - c_j| - (r_i + r_j),
\end{equation*}
where $c_i = \boldsymbol{t}_0^i \cdot \hat{\boldsymbol{a}}$ and $r_i = \sum_{d \in \{x,y,z\}} s_{0,d}^i |\boldsymbol{n}_i^{d} \cdot \hat{\boldsymbol{a}}|$.
The distance between two OBBs is then:
\begin{equation}
\label{eq:contact_distance}
d(P_i, P_j) = \max_{\mathbf{a} \in \mathcal{A}} d_{\mathbf{a}}, 
\end{equation}
whose positive values indicate separation, zero indicates contact, and negative values indicate overlap. This differentiable distance metric enables efficient enforcement of contact constraints during optimization (see Sec.~\ref{sec:optimization}).

%% file: figs/overview.tex
\begin{figure}[t]
    \centering
    \includegraphics[width=\linewidth,trim={0cm 22.7cm 0cm 0cm}]{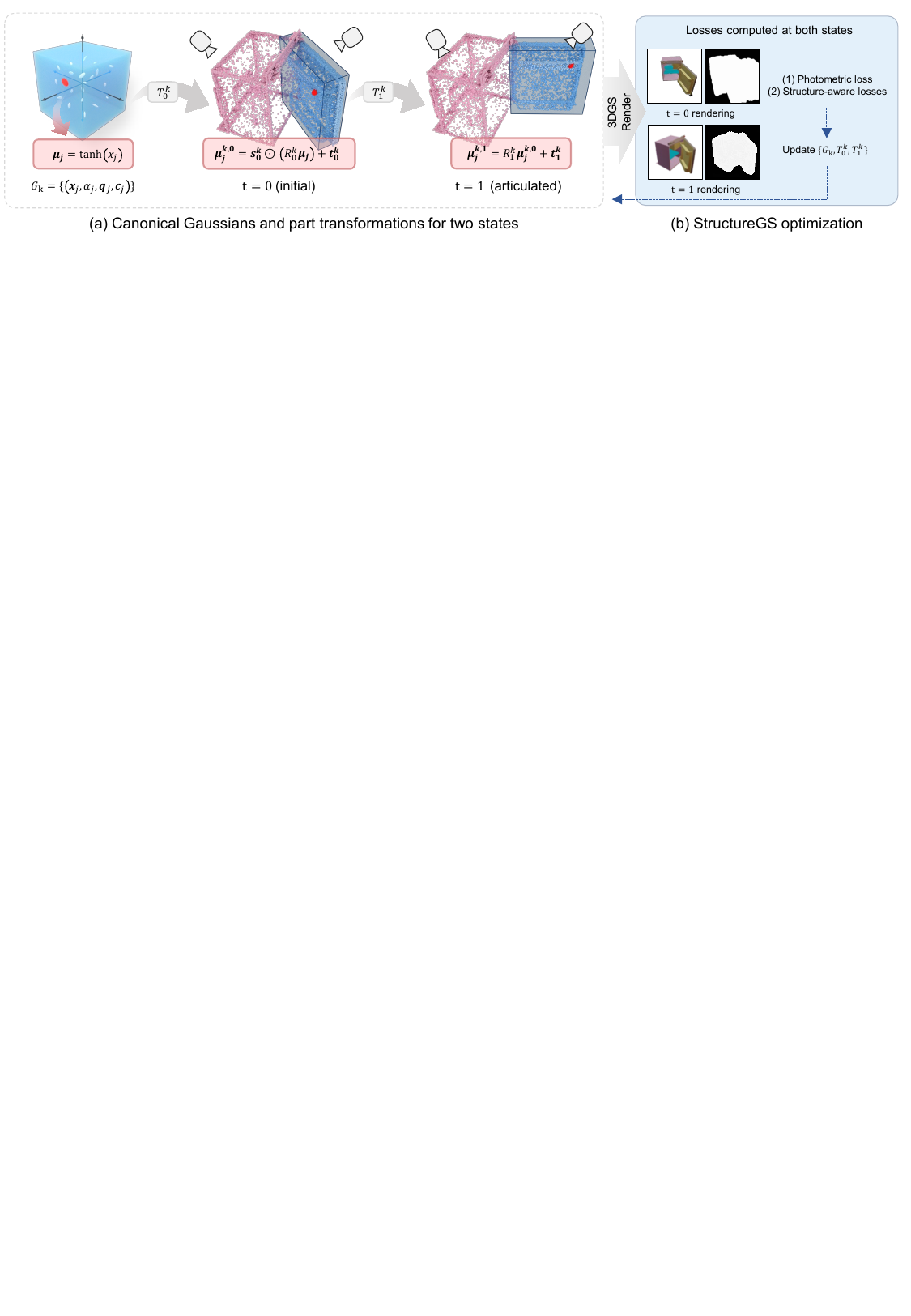}
    \caption{Overview of StructureGS. (a) Each part $P_k$ is represented by canonical Gaussians $G_k$, which are transformed to two states: an initial state ($t=0$) and an articulated state ($t=1$). (b) For each state, the transformed Gaussians are rendered with the 3DGS renderer, and photometric and structure-aware losses are computed to jointly optimize $G_k$, $T_0^k$, and $T_1^k$.}
    \label{fig:overview}
    
\end{figure}

%% file: figs/sat.tex
\begin{wrapfigure}{r}{0.23\textwidth}
    \centering\vspace{-20pt}
\includegraphics[width=\linewidth,trim={0cm 23cm 13cm 1cm}]{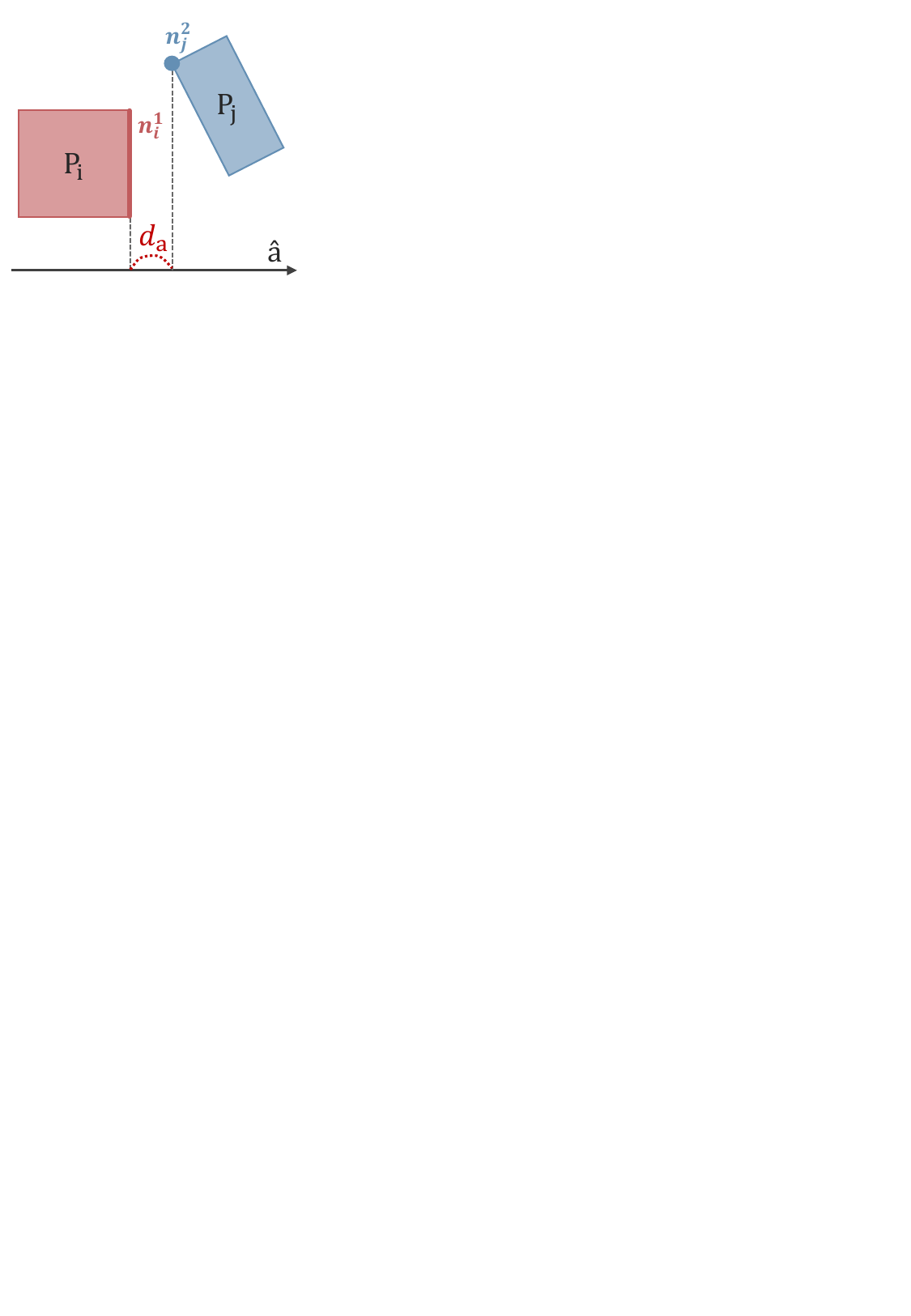}
\vspace{-14pt}
\end{wrapfigure}

%% file: sec/4_Optimization.tex
\section{Optimization with Structural Guidance} 
\label{sec:optimization}

For articulated object reconstruction with meaningful part decomposition, in addition to the photometric loss, we introduce two differentiable losses for structural guidance: \textit{part fitting loss} for spatial coherence of parts and \textit{part contact loss} for structural connectivity of parts (see \Fig{overview}). 
    
\subsection{Loss Functions}

\paragraph{Photometric image loss}
Following 3D Gaussian Splatting~\cite{kerbl20233d}, we employ the standard photometric image loss:
\begin{equation}
\mathcal{L}_{\text{image}} = (1 - \lambda_{\text{ssim}}) \mathcal{L}_1 + \lambda_{\text{ssim}} \mathcal{L}_{\text{D-SSIM}},
\end{equation}
where $\mathcal{L}_1$ measures the L1 distance between the rendered and ground-truth images, and $\mathcal{L}_{\text{D-SSIM}}$ is the structural dissimilarity term.

\input{figs/fitting_loss}
\paragraph{Part fitting loss}
We enforce spatially coherent parts by ensuring each bounding box tightly encloses its part geometry with minimal unused interior volume. To quantify this constraint, we first generate structure images that visualize each part's spatial extent: we uniformly sample points within the canonical cube $[-1, 1]^3$, transform them to the world space using $T_0^k$ for $t=0$ (or additionally apply $T_1^k$ for $t=1$), and render them as white pixels on a black background.
The rendered structure image is a binary image that represents the spatial occupancy of each part's bounding box, where 1 (white) indicates occupied regions.
Based on the structure images, we formulate the part fitting loss with two complementary objectives:
\begin{equation*}
\mathcal{L}_{\text{fit}} = \lambda_{\text{coverage}} \mathcal{L}_{\text{coverage}} + \lambda_{\text{margin}} \mathcal{L}_{\text{margin}}.
\end{equation*}

The coverage loss $\mathcal{L}_{\text{coverage}}$ ensures that the union of all part structure images fully covers the object region in the rendered image:
\begin{equation*}
\mathcal{L}_{\text{coverage}} = \| (1 - \mathbf{I}_{\text{struct}}^{\text{all}}) \odot \mathbf{M} \|_1,
\end{equation*}
where $\mathbf{I}_{\text{struct}}^{\text{all}}$ is the union of structure images rendered from OBBs of all parts, $\mathbf{M}$ is the object mask in the rendered image, and $\odot$ denotes element-wise multiplication. This term penalizes regions of the object not covered by any OBB. 

Minimizing $\mathcal{L}_{\text{coverage}}$ alone may cause OBBs to expand excessively beyond object boundaries. We therefore introduce a margin loss that penalizes regions of each part's structure image that lie outside the object mask:
\begin{equation}
\mathcal{L}_{\text{margin}} = \frac{1}{K} \sum_{k=0}^{K-1} \| \mathbf{I}_{\text{struct}}^k \odot (1 - \mathbf{M}) \|_1,
\label{eq:margin}
\end{equation}
where $\mathbf{I}_{\text{struct}}^k$ denotes the structure image of part $P_k$. This term encourages tight fitting of OBBs by penalizing the unused void space within each bounding box. 
For concave parts, reaching the tightest feasible bound does not necessarily drive the margin loss to zero, as unavoidable empty regions may remain within their corresponding OBBs.
\Fig{fitting_loss} provides a visual illustration of the part fitting loss. 

\paragraph{Part contact loss} 
To enforce connectivity, we penalize separation between adjacent parts using the distance $d(\cdot, \cdot)$ computed by Eq.~{\ref{eq:contact_distance}}:
\begin{equation}
\mathcal{L}_{\text{contact}} = \lambda_{\text{contact}}\sum_{(i,j) \in \mathcal{E}} \max(0, d(P_i, P_j))^2 ,
\end{equation}
\begin{wrapfigure}{r}{0.4\textwidth} 
  \centering
  \vspace{-25pt} 
    \includegraphics[width=0.4\textwidth,trim=0cm 8.5cm 9cm 0cm, clip]{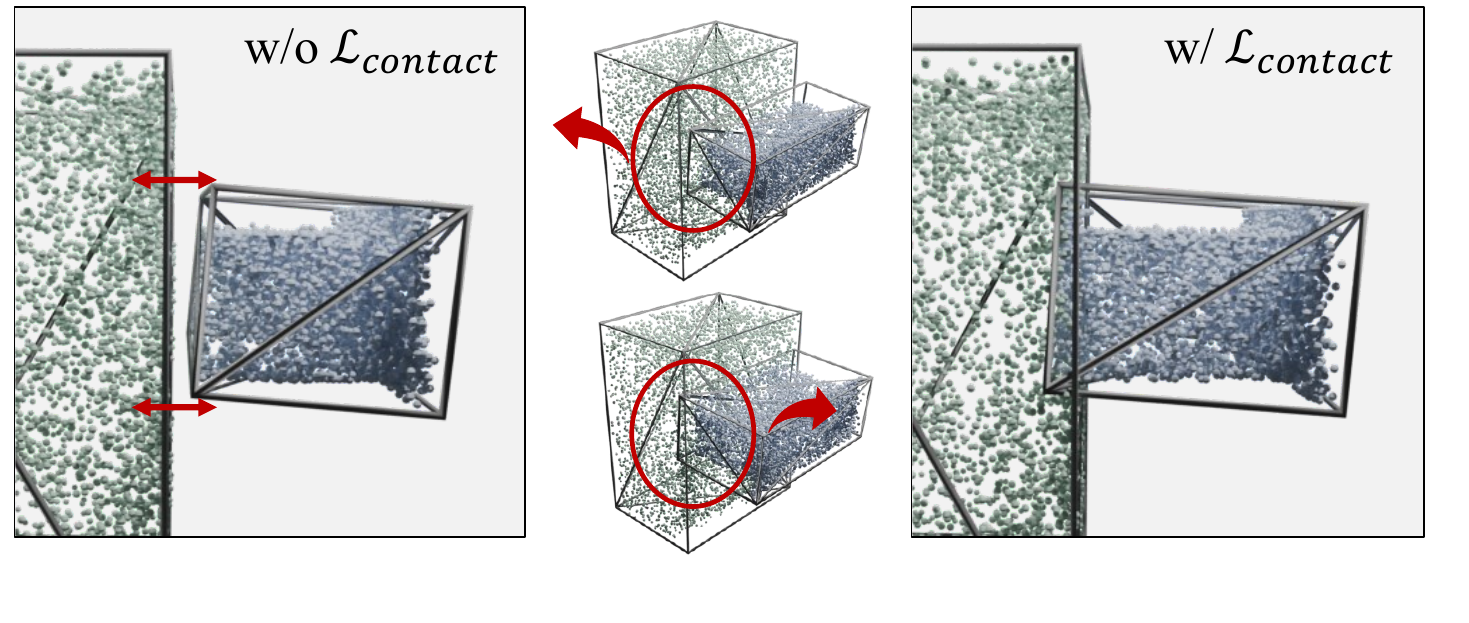}
  \vspace{-35pt} 
\end{wrapfigure}
where $\mathcal{E}$ denotes the set of adjacent pairs, formed between the static base part $P_0$ and each movable part $P_k$ ($k \geq 1$). 
Since OBBs encode part structures rather than exact occupied volumes, valid articulated layouts can involve OBB overlap, as in a drawer nested inside a cabinet body. Accordingly, the $\max$ operator penalizes only separation, not overlap.

\paragraph{Extent regularization}
To encourage the OBB scale components to adapt to the underlying part geometry, we introduce an axis-weighted scale regularization:
\begin{equation}
\mathcal{L}_{\text{ext}} = -\lambda_{\text{ext}} \sum_{k=0}^{K-1} \sum_{i=1}^{3} w_k^{(i)} \log(s_{0,i}^k),
\end{equation}
where $s_{0,i}^k$ is the $i$-th scale component of the $k$-th OBB, and $\mathbf{w}_k = \text{softmax}(\boldsymbol{\beta}_k)$ are per-axis weights 
with learnable parameters $\boldsymbol{\beta}_k \in \mathbb{R}^3$. 
The softmax weights act as a differentiable selector over the OBB scale components, 
emphasizing scale directions whose expansion does not increase the overall objective.
Consequently, the OBB expands primarily along under-covered directions rather than scaling uniformly.

Combining structure-aware and photometric image losses across temporal states $t \in \{0, 1\}$, along with extent regularization, our final optimization objective is:
\begin{equation}
\mathcal{L}_{\text{total}} = \sum_{t \in \{0,1\}} \left( \mathcal{L}_{\text{image}}^t + \mathcal{L}_{\text{fit}}^t + \mathcal{L}_{\text{contact}}^t \right) + \mathcal{L}_{\text{ext}}.
\end{equation}

\begin{figure*}[t]
\begin{center}
    \centering
    \includegraphics[width=\textwidth,trim=0cm 14cm 2cm 0cm, clip]{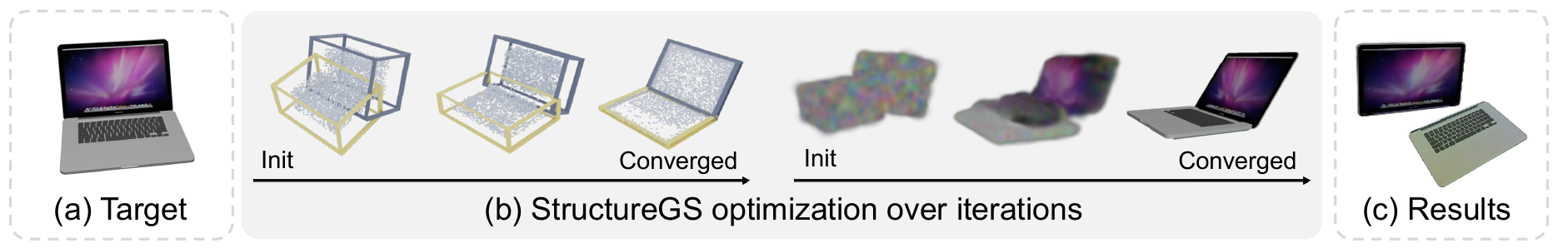}
\end{center}
    \caption{ 
    Structure-aware optimization over iterations. Part-wise 3D Gaussian splats constrained by OBBs progressively align with their corresponding object parts.
    }
    \label{fig:iteration}
\end{figure*}

\subsection{Optimization Process}
We initialize Gaussians by randomly sampling 10,000 primitives per part within the canonical space $[-1,1]^3$. For OBB parameters, we initialize the boxes using coarse point clouds estimated for the two articulation states with VGGT~\cite{wang2025vggt}. We use nearest-neighbor distances between the two states to separate static and moving regions, cluster the moving points into dynamic parts with DBSCAN~\cite{ester1996density}, and fit an OBB to each resulting part to initialize $T_0^k$. The articulation transforms $T_1^k$ are initialized as identity.
The initialization provides only a coarse structural cue and does not require accurate part segmentation. The proposed structure-aware losses subsequently optimize both part geometry and motion, making the method tolerant to imperfect initial bounding boxes. We provide detailed initialization procedures and robustness analysis in the supplementary material.

Optimization proceeds for 10,000 iterations using Adam optimizer~\cite{kingma2017adammethodstochasticoptimization} with progressive scheduling. Structure-aware losses ($\mathcal{L}_{\text{fit}}$, $\mathcal{L}_{\text{contact}}$, $\mathcal{L}_{\text{ext}}$) are active from the start to establish part decomposition. 
We introduce  $\mathcal{L}_{\text{image}}$ at iteration 400 and disable $\mathcal{L}_{\text{contact}}$ and $\mathcal{L}_{\text{ext}}$ at iteration 1500, once the part structures are sufficiently established.
During optimization, we apply standard 3DGS densification and pruning.
More details on optimization schedules, hyperparameters, and architectural details are provided in the supplementary material.

In essence, we reparameterize each part as a canonical Gaussian set in the unit cube centered at the origin. Before rendering, the canonical Gaussians are first transformed to the initial state by the OBB transform $T_0^k$, and then mapped to the articulated state by the articulation transform $T_1^k$. This differentiable parameterization allows photometric and structure-aware losses to jointly update the Gaussians, OBB parameters, and articulation parameters, guiding the reconstruction toward coherent articulated parts.

\Fig{iteration} visualizes the optimization process. The left columns show part-wise bounding boxes and their associated Gaussian primitive sets, while the right columns show rendered Gaussians. As optimization progresses, the structure-aware losses guide the Gaussian sets toward separated parts with cleaner boundaries.

%% file: figs/fitting_loss.tex
\begin{figure}[t]
    \centering
    \includegraphics[trim={0cm 24cm 0cm 0cm},width=\linewidth]{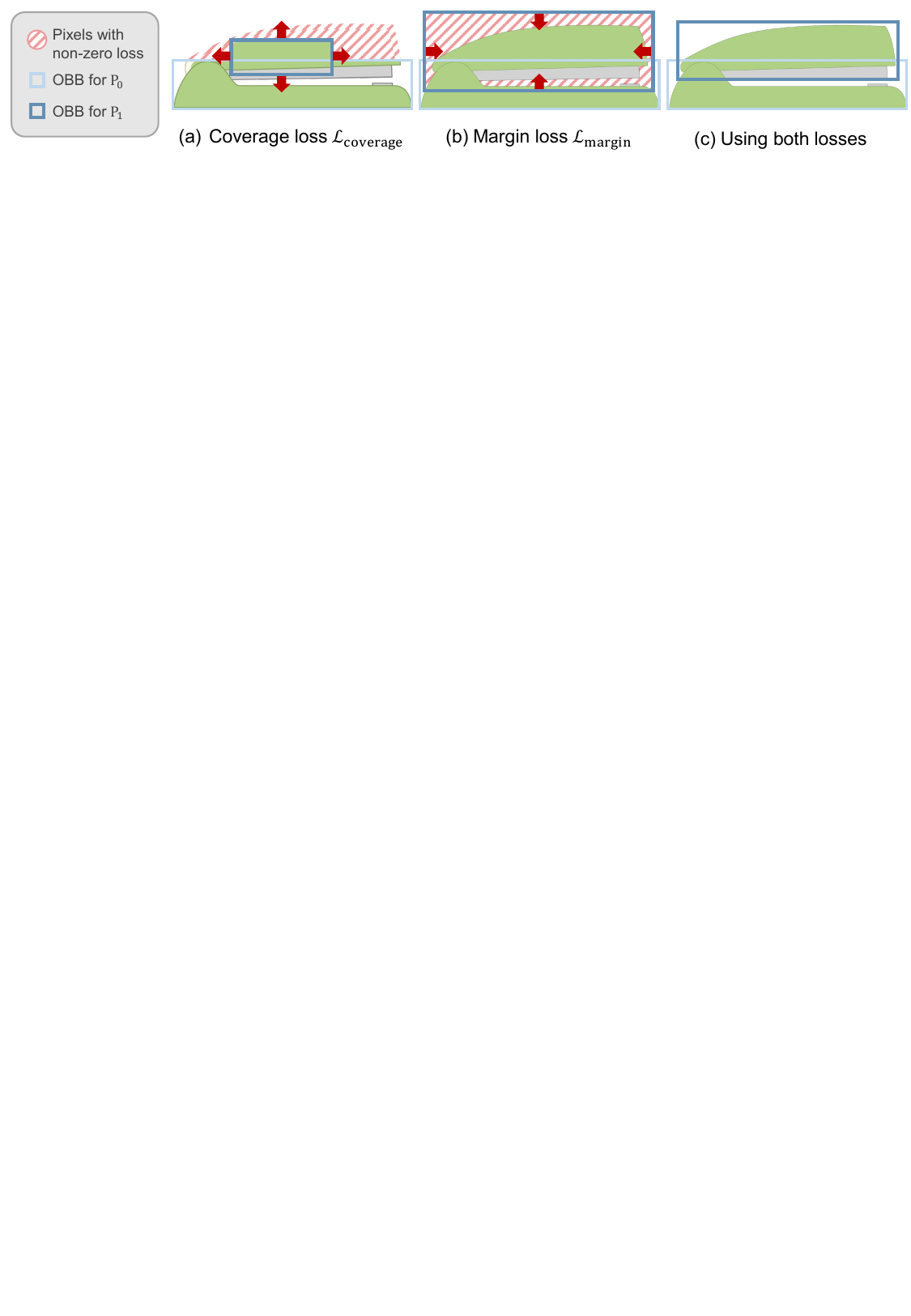}
    \caption{Illustration of part fitting loss. (a) $\mathrm{P}_1$'s OBB is fully inside the part (margin loss = 0), but does not cover the entire part, incurring coverage loss. (b) $\mathrm{P}_1$'s OBB fully covers the part (coverage loss = 0), but includes regions outside the part, incurring margin loss. (c) Optimizing both losses yields a tight OBB.
    } 
    \label{fig:fitting_loss}
    
\end{figure}

%% file: sec/5_Experiments.tex
\section{Experiments}
\label{sec:experiments}


\subsection{Setup}

\paragraph{Benchmark datasets}
We evaluate our method on two primary articulated object benchmarks. The PARIS dataset~\cite{liu2023paris} comprises 10 synthetic and 2 real-world objects, each consisting of a static base and a single movable part. The DTA-Multi dataset~\cite{weng2024neural} contains 2 synthetic multi-part objects, each consisting of a static base and multiple movable parts. For both datasets, objects are captured in two distinct articulation states, each with 100 multi-view RGB images at an $800 \times 800$ resolution from cameras uniformly distributed over a hemisphere. To assess robustness under limited observations, we additionally introduce a sparse-view setting using only 10 uniformly sampled views per state from PARIS. Additional scalability experiments on more complex articulated objects with 4--7 parts are provided in the supplementary material.


\paragraph{Real-world data}
To demonstrate practical usefulness of our method, we introduce a custom real-world dataset of four articulated objects (kettle, stand, carrier, and hole punch) that exhibit challenging features: textureless surfaces, non-box-like geometries, and subtle articulations. As a practical setup, we utilize only 10 images per state captured via a consumer smartphone. Camera poses and object masks are obtained using off-the-shelf tools, COLMAP~\cite{schoenberger2016sfm} and SAM~\cite{kirillov2023segment}, respectively, so our evaluation reflects the estimation noise encountered in practical settings.



\paragraph{Baselines} 
We compare our method against three representative baselines using their official implementations: PARIS~\cite{liu2023paris}, which explicitly decomposes part-wise neural radiance fields; ArticulatedGS~\cite{guo2025articulatedgs}, which discovers part structures via deformation-aware clustering on canonical 3D Gaussians; and ScrewSplat~\cite{kim2025screwsplat}, which models continuous screw motions using soft Gaussian assignments. For a fair comparison, we adapt ScrewSplat to the two-state setting according to its original protocol, evaluating multiple parsimony weights per scene to report the best achievable result.



\paragraph{Evaluation metrics} 
We comprehensively evaluate our method across three aspects: geometry, kinematics, and rendering. Following~\cite{liu2023paris, guo2025articulatedgs}, geometric accuracy is measured via the Chamfer-$L_1$ distance (CD) using $10,000$ sampled points, reported for the whole object (CD-w), static base (CD-s), and movable part (CD-m), all scaled by $10^3$. For kinematics, we evaluate joint axis accuracy using angular error (in degrees) for all joints and position error specifically for revolute joints. Motion magnitude is assessed via geodesic and translation errors for revolute and prismatic motions, respectively. Finally, novel view synthesis is evaluated using PSNR, SSIM, and LPIPS.

\begin{figure*}[t]
\begin{center}
    \centering
    \includegraphics[width=\linewidth,trim=0cm 8.5cm 11cm 0cm, clip]{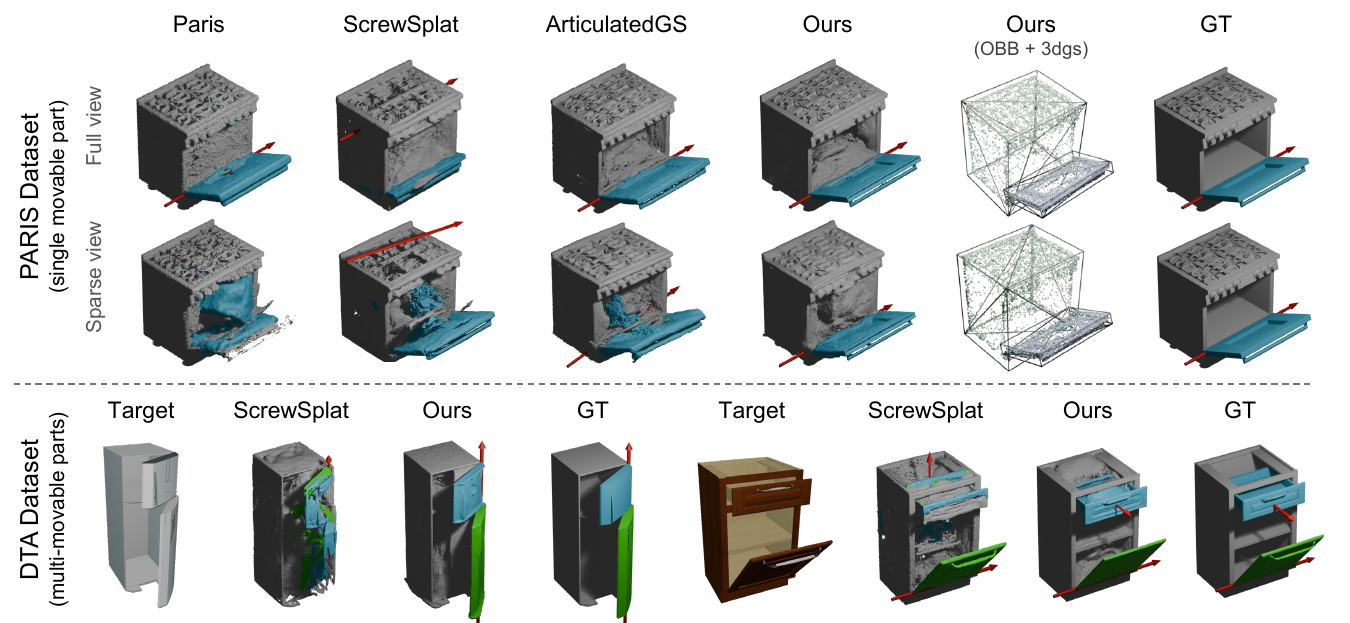}
\end{center}
    \caption{Qualitative comparison on PARIS and DTA datasets under full- and sparse-view settings. Reconstructed geometry, fitted OBBs with per-part 3DGS, and estimated joint axes (red) are shown; gray denotes static parts, while blue and green indicate movable parts. Additional results are provided in the supplementary material.
    }
    \label{fig:quali}
\end{figure*}

\begin{figure}[t]
\begin{center}
    \includegraphics[width=1.\linewidth,trim=0cm 11cm 0cm 0cm, clip]{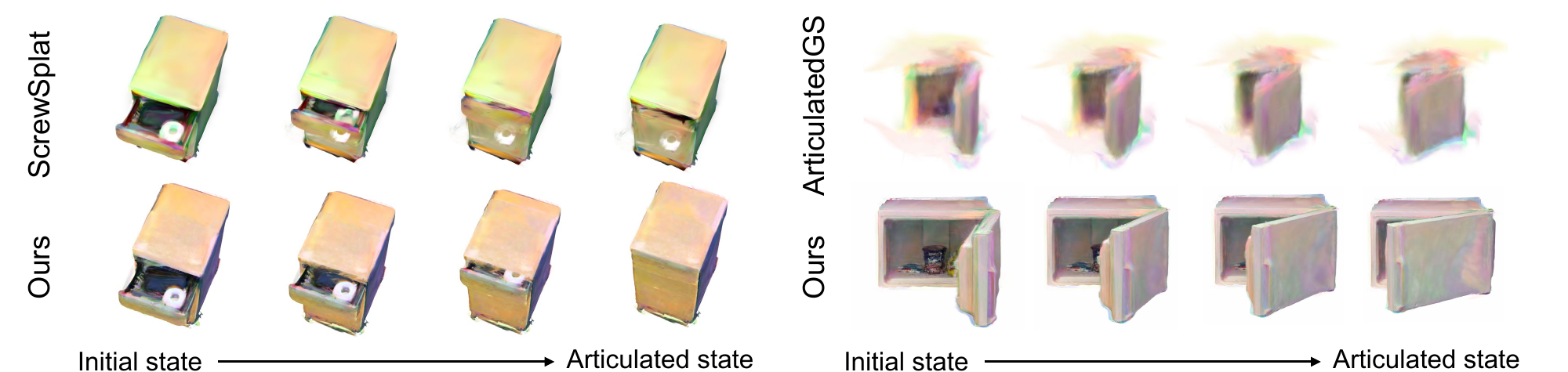}
\end{center}
    \caption{ 
  Articulation animation on PARIS real data from initial to articulated states.   
    }
    \label{fig:animation}
\end{figure}

\begin{table}[t]
\centering
\caption{Quantitative results on the PARIS dataset under full (100 views) and sparse (10 views) settings. Scores are averaged over 8 revolute and 2 prismatic scenes for each joint type; per-scene results are provided in the supplementary material. Angle error is in degrees (°); CD is scaled by ×1000. }
\label{tbl:paris_quant}
\small
\resizebox{\columnwidth}{!}{
\begin{tabular}{l @{\hspace{10pt}}  l @{\hspace{10pt}} l @{\hspace{10pt}}  c c c >{\columncolor{ourscolor}}c @{\hspace{10pt}}  c c c >{\columncolor{ourscolor}}c }
\toprule
\multirow{2}{*}{Group}
& \multirow{2}{*}{Metric} & \multirow{2}{*}{Type}
& \multicolumn{4}{c}{PARIS full views (100)}
& \multicolumn{4}{c}{PARIS sparse views (10)} \\ 
\cmidrule(lr){4-7}\cmidrule(lr){8-11} 
& & 
& PARIS & ScrewSplat & ArticulatedGS & Ours
& PARIS & ScrewSplat & ArticulatedGS & Ours \\ 
\midrule
\multirow{4}{*}{Motion}
& Ang Err
& All
& 5.635 & 7.787 & 0.091 & \textbf{0.088} & 32.840 & 11.537 & 8.553 & \textbf{0.701} \\
\cmidrule{2-11}
& \multirow{1}{*}{Pos Err} 
 &   Revolute   & 0.385 & 0.139 & 0.018 & \textbf{0.001} & 0.162 & 0.181 & 0.126 & \textbf{0.007}  \\
\cmidrule{2-11}

& \multirow{2}{*}{Geo Dist}
&   Revolute   & 85.52 & 32.20 & 0.761 & \textbf{0.066} & 71.84 & 39.96 & 12.89 & \textbf{1.305}  \\
& &   Prismatic  & 1.062 & 0.200 & 0.200 &\textbf{ 0.001 }& 0.14 & 0.401 & 0.270 & \textbf{0.042}  \\
\midrule
\multirow{3}{*}{Geometry}
& CD-s
& All
& 5.528 & 38.948 & 2.119 & \textbf{1.558} & 157.46 & 36.521 & 2.044 & \textbf{1.957} \\
\cmidrule{2-11}

& CD-m
& All
& 67.935 & 32.202 & 1.585 & \textbf{1.400} & 164.93 & 133.50 & 62.700 & \textbf{1.982} \\
\cmidrule{2-11}

& CD-w
& All
& 5.895 & 13.935 & 1.908 & \textbf{0.919} & 84.632 & 31.892 & 19.847 & \textbf{1.320} \\
\midrule
\multirow{3}{*}{Photometry}
& PSNR $\uparrow$  & All &  29.324 & 31.377  & 37.336  & \textbf{43.39} & 22.363 & 28.135 & 23.320 & \textbf{33.264} \\
& SSIM $\uparrow$  & All & 0.948  & 0.965  & 0.985  & \textbf{0.995} & 0.878 & 0.952 & 0.910 & \textbf{0.969} \\
& LPIPS $\downarrow$  & All & 0.097  & 0.046  & 0.039  & \textbf{0.014} & 0.160 & 0.064 & 0.095 & \textbf{0.052} \\

\bottomrule
\end{tabular}}
\vspace{-10pt}
\end{table} 

\subsection{Comparisons}
Our method achieves state-of-the-art performance in both dense- and sparse-view settings, with consistent improvements across geometry, motion, and photometric metrics. Notably, baselines exhibit inconsistency in metrics: a low whole-object Chamfer Distance often obscures severe part-level geometric errors (\Tbl{paris_quant}) and incorrect motion estimates (\Fig{quali}). This suggests that photometric-only optimization can converge to physically implausible states that satisfy visual observations but misrepresent true articulation. Moreover, this ambiguity between geometry and motion is amplified in under-constrained sparse-view settings.

Consequently, during animation, baseline reconstructions suffer from ghosting artifacts and blurry boundaries caused by flawed part decomposition (\Fig{animation}). In contrast, our structure-aware formulation jointly regularizes part decomposition and motion, ensuring physically valid articulation and robust reconstructions regardless of view density.

We further validate our method's scalability on the DTA dataset, which features more complex, multi-part kinematic structures. As shown in \Tbl{dta_quanti} and \Fig{quali}, our approach maintains high-quality reconstructions and lower errors across both scenes. While ScrewSplat achieves a marginally lower error for one joint in the \textit{Storage} scene, it suffers from significant geometric and kinematic inaccuracies elsewhere. Specifically, it fails to decouple the two movable parts, merging them into a single component and propagating errors that degrade the overall reconstruction. 
Further comparisons with ScrewSplat on ArtGS-Multi dataset in the supplementary material show consistent gains on more complex 4--7 part articulated objects.

Finally, in the full-view setting, our method converges in approximately 13 minutes per scene, achieving comparable efficiency to ArticulatedGS (13 min) and ScrewSplat (15 min), while being substantially faster than PARIS (40 min).

\begin{table}[t]
\centering
\caption{Quantitative results on the DTA dataset, reported per part. Superscripts $^{\dagger}$ and $^{\ddagger}$ denote part 0 and part 1, respectively. F denotes failure; `-' indicates not applicable due to prismatic geometry. }
\label{tbl:dta_quanti}
\small
\resizebox{\columnwidth}{!}{
\begin{tabular}{l l c c c c c c c c c c }
\toprule
Object & Method & Ang Err$^{\dagger}$ & Ang Err$^{\ddagger}$ & Pos Err$^{\dagger}$ & Pos Err$^{\ddagger}$ & Geo Dist$^{\dagger}$ & Geo Dist$^{\ddagger}$ & CD-s & CD-m$^{\dagger}$ & CD-m$^{\ddagger}$ & CD-w  \\
\midrule

\multirow{2}{*}{Fridge} 
& ScrewSplat  & 15.12 & 0.347 & 0.024 &0.007 &9.640 &15.12 & 25.43& 321.01& 34.93& 8.369\\
&\cellcolor{ourscolor} Ours &\cellcolor{ourscolor} \textbf{0.445} &\cellcolor{ourscolor} \textbf{0.119}& \cellcolor{ourscolor}\textbf{ 0.003}& \cellcolor{ourscolor}\textbf{0.002}& \cellcolor{ourscolor}\textbf{1.301}&\cellcolor{ourscolor} \textbf{0.217 }&\cellcolor{ourscolor} \textbf{5.888} & \cellcolor{ourscolor}\textbf{0.736} & \cellcolor{ourscolor}\textbf{1.078} & \cellcolor{ourscolor}\textbf{4.699}\\
\midrule

\multirow{2}{*}{Storage} 
& ScrewSplat & \textbf{ 0.052}& 89.86 & \textbf{0.003} &  -& 60.30 & -&136.1 & 180.9& 299.5& 38.15\\
& \cellcolor{ourscolor}Ours       & \cellcolor{ourscolor}2.165 & \cellcolor{ourscolor} \textbf{0.542} &  \cellcolor{ourscolor}0.017&\cellcolor{ourscolor} - &\cellcolor{ourscolor}  \textbf{0.292} & \cellcolor{ourscolor}-& \cellcolor{ourscolor} \textbf{3.336}  & \cellcolor{ourscolor} \textbf{0.553} &\cellcolor{ourscolor}  \textbf{4.184} & \cellcolor{ourscolor} \textbf{3.238} \\

\bottomrule
\end{tabular}}
\end{table}

\subsection{Real-world evaluation}
\begin{figure*}[t]
\begin{center}
    \centering
    \includegraphics[width=1.\linewidth,trim=0cm 9cm 0cm 0cm, clip]{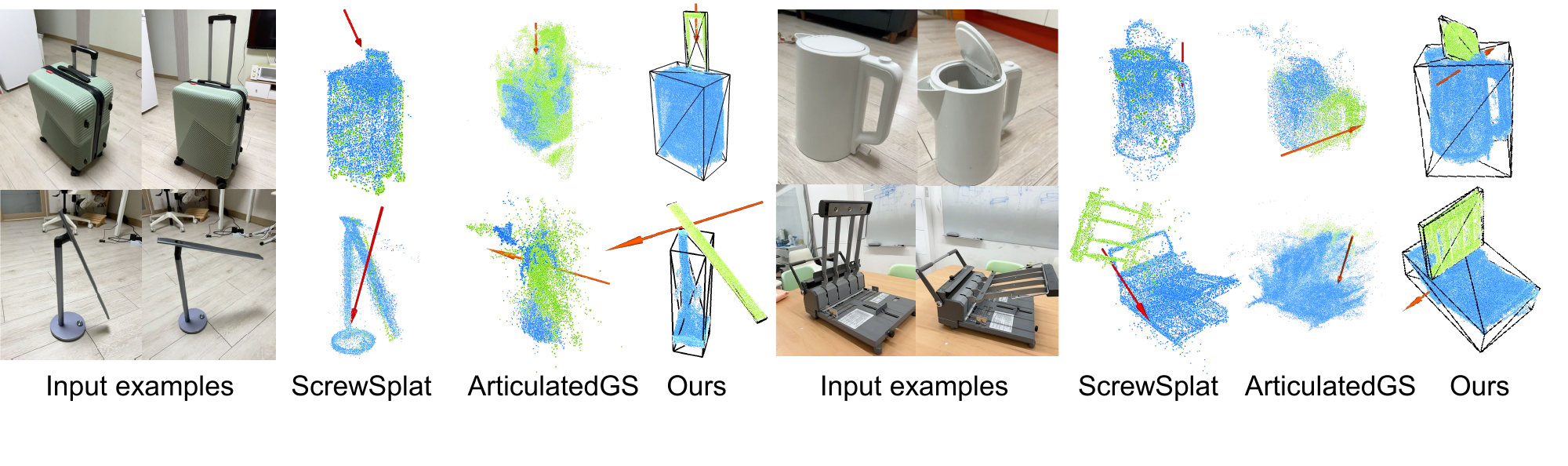}
\end{center}
    \caption{Reconstruction of real-world articulated objects using 10 views. 
    Blue/green Gaussian splats denote static/dynamic parts, respectively. 
    Joint axes and estimated OBBs of our method are shown as red arrows and black wireframes, respectively.}
    \label{fig:real}
\end{figure*}

We further evaluate our method's generalization under sparse and noisy real-world conditions using only 10 views per state. As shown in \Fig{real}, photometric-only baselines struggle to recover geometry and articulation, producing scattered Gaussians and inaccurate joint estimates. In contrast, our method reconstructs coherent part-level 3DGS and accurate kinematics by leveraging tightly fitted OBBs that properly bound each semantic part. These results demonstrate that our explicit structural guidance effectively generalizes to diverse real-world articulated objects. Additional animations are available in the supplementary material.







\begin{figure*}[t]
  \centering
  \begin{minipage}[c]{0.52\textwidth}
    \centering
    \includegraphics[width=0.95\linewidth]{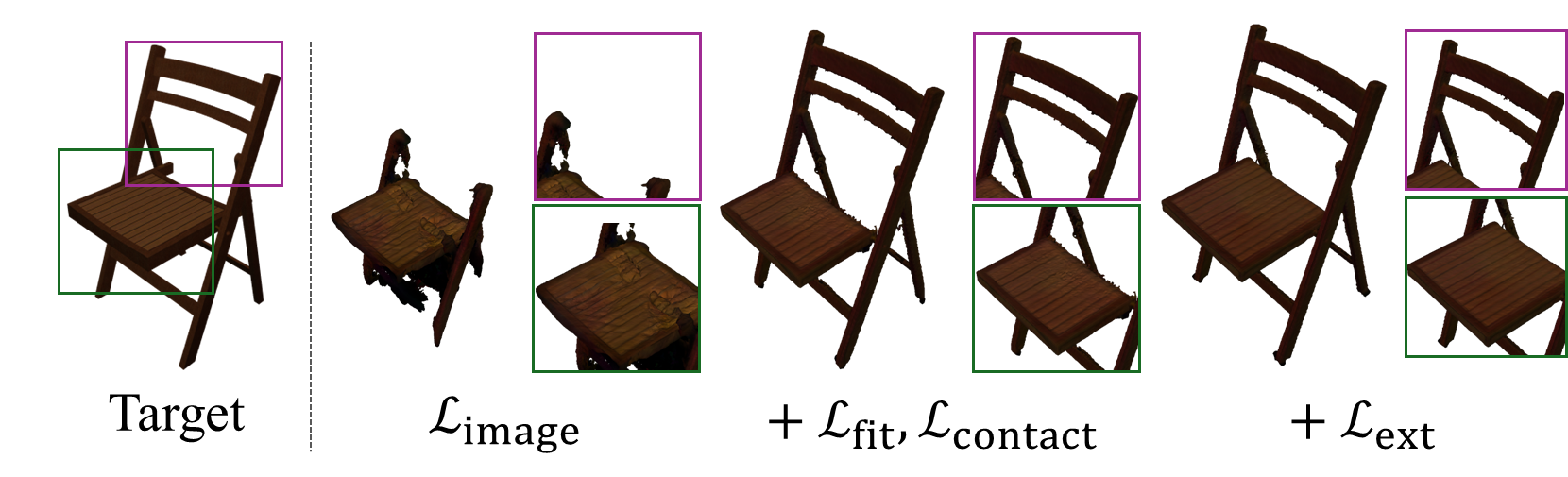}
  \end{minipage}
  \hfill 
  \begin{minipage}[c]{0.45\textwidth}
    \centering
    \resizebox{\linewidth}{!}{
      \begin{tabular}{l cccc ccc}
        \toprule          
        Loss & Ang & Pos & GD (r) & GD (p) & CD-s & CD-m & CD-w \\
        \midrule
        $\mathcal{L}_{\text{image}}$ & 5.60 & 0.11 & 7.65 & 0.01 & 6.08 & 31.8 & 3.90 \\
        + $\mathcal{L}_{\text{fit, con}}$ & 1.71 & 0.00 & 0.06 & 0.00 & 1.89 & 31.4 & 0.98 \\
        + $\mathcal{L}_{\text{ext}}$ & 0.08 & 0.00 & 0.06 & 0.00 & 1.56 & 1.40 & 0.91 \\
        \bottomrule
      \end{tabular}
    }
  \end{minipage}

  \caption{Ablation on structural losses. Left: Visual comparison of the effect of each loss term on reconstruction quality. Right: Quantitative results for each loss configuration.}
  \label{fig:combined_ablation}
\end{figure*}

\subsection{Discussion}
\paragraph{Loss function ablation}
To validate our structural losses, we conduct an ablation study on the PARIS dataset (\Fig{combined_ablation}). Optimizing solely with the photometric loss ($\mathcal{L}_{\text{image}}$) yields severe geometry and kinematic errors due to unresolved shape-motion ambiguity. Introducing the fitting and contact losses ($\mathcal{L}_{\text{fit}}$, $\mathcal{L}_{\text{contact}}$)  reduces both geometric distortion and joint inaccuracies, enabling coherent part decomposition and stable kinematics. Finally, adding the extent regularization ($\mathcal{L}_{\text{ext}}$) rectifies under-estimated bounding volumes, stabilizing part extents and further decreasing geometric error. 
Notably, the large CD-m reduction after adding $\mathcal{L}_{\text{ext}}$ mainly comes from the Blade scene. There, the regularizer helps the OBB recover the full movable-part extent, rather than fitting only to the exposed tip where silhouette changes are most dominant.

\begin{figure*}[t]
    \centering
        \includegraphics[width=0.9\linewidth,trim=0cm 12cm 3.5cm 0cm, clip]{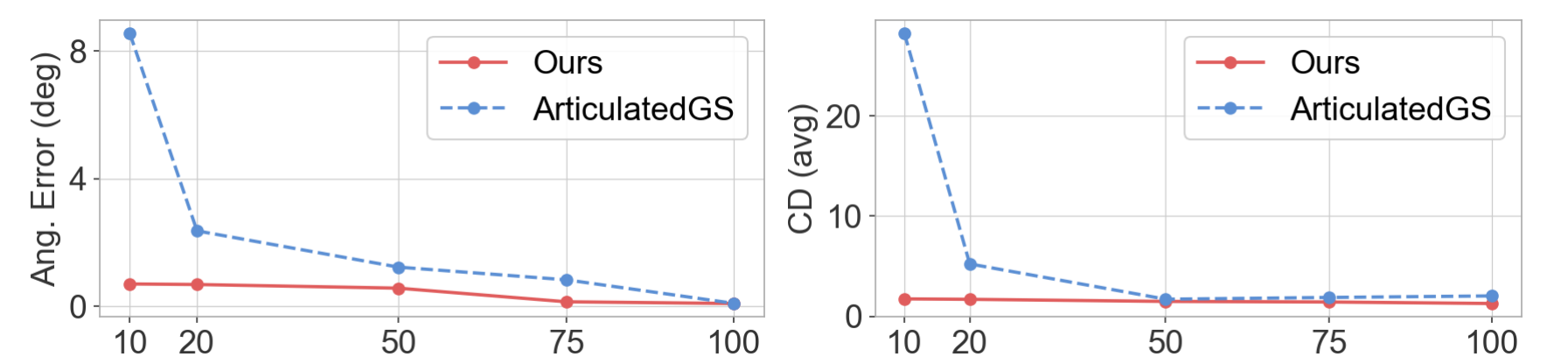}
    \caption{ Robustness to sparse observations. Angular error and Chamfer distance on the PARIS benchmark under varying numbers of training views. Our method remains stable as views become sparse, while ArticulatedGS~\cite{guo2025articulatedgs} degrades sharply.}
    \label{fig:sp_robustness} 
\end{figure*}
\paragraph{Robustness to view sparsity}
We evaluate the robustness of our model to view sparsity on the PARIS benchmark dataset~\cite{liu2023paris}. We optimize models using 10
to 100 uniformly subsampled training views and compare geometric and motion errors against ArticulatedGS~\cite{guo2025articulatedgs}. 
\Fig{sp_robustness} shows that, as observations become sparser, ArticulatedGS exhibits a sharp increase in both
angular error and Chamfer distance, indicating severe shape-motion ambiguity under limited photometric supervision. In contrast, our method maintains stable performance across view counts, demonstrating that OBB-based structural
guidance effectively constrains the solution space when visual evidence is limited.
 
\paragraph{Non-OBB-shaped objects} 
Although our structure representation uses OBBs, StructureGS is not constrained to box-like shapes. 
By jointly optimizing each OBB with its enclosed 3D Gaussians, the OBB provides a compact proxy for the coarse part structure, while the Gaussians recover fine-grained geometry.
For instance, the static part of the \textit{foldchair} scene (PARIS benchmark)
\begin{wrapfigure}{r}{0.22\textwidth} 
  \centering
  \vspace{-20pt} 
    \includegraphics[width=1.\linewidth,trim=0cm 13cm 28cm 0.5cm, clip]{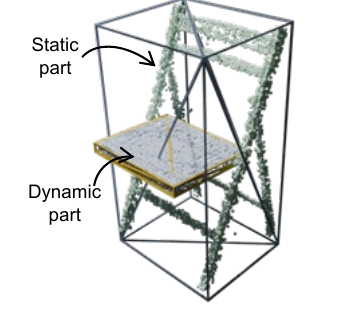}
  \vspace{-35pt} 
\end{wrapfigure} 
 consists of extremely thin structures. To quantify this non-cuboidal nature, 
we define an empty space ratio $r = 1 - V_{\text{mesh}}/V_{\text{OBB}}$. For the \textit{foldchair}, we obtain $r = 0.97$, meaning the bounding volume contains $97\%$ empty space. As illustrated in our qualitative results, even in such extreme cases, the OBB-constrained Gaussians successfully recover intricate geometric details. 



%% file: sec/6_Conclusion.tex
\section{Conclusion}
\label{sec:conclusion}

In this work, we presented StructureGS, a framework for articulated object reconstruction that achieves clean part decomposition through structure-aware guidance. By representing parts with oriented bounding boxes and enforcing spatial coherence and structural connectivity constraints, our method produces well-defined part boundaries and geometrically coherent structures. Extensive experiments demonstrate that StructureGS outperforms existing methods in both geometric quality and articulation accuracy.
\begin{figure*}[t]
\begin{center}
    \centering
    \includegraphics[width=0.75\linewidth,trim=0cm 15cm 17cm 0cm, clip]{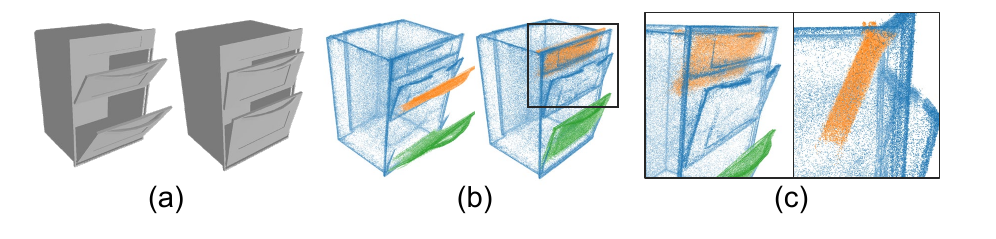}
\end{center}
    \caption{Failure case. (a) Target object in the initial and articulated states. (b) Reconstructed 3DGS in the corresponding states, with parts color-coded. (c) Zoomed-in views showing part collapse in the articulated state.}
    \label{fig:failure}
\end{figure*}

\paragraph{Limitation}
Although our OBB-based formulation is more robust than photometric-only optimization, it can still fail when articulated motion induces only weak visual changes. For example, in the oven case in \Fig{failure}, the upper moving part provides limited silhouette and appearance cues, causing its geometry to be absorbed by the nearby static region. As a result, the dynamic part can collapse into the static body, producing implausible geometry and motion. Incorporating higher-level semantic features such as DINO~\cite{oquab2024dinov2learningrobustvisual} could further improve robustness in such cases.

%% file: sec/X_suppl.tex
In this supplementary material, we provide additional details on implementation, robustness analysis for OBB initialization, and further experimental results.


\section{Implementation Details}
\subsection{Training configuration}
We optimize our model using the Adam optimizer~\cite{kingma2017adammethodstochasticoptimization} for 10,000 iterations. For optimizing 3D Gaussian Splatting (3DGS) parameters, we use the same learning rates as in the original 3DGS framework~\cite{kerbl20233d}, except for the Gaussian means, which use a learning rate of 1.6e-2. 
The canonical-to-world transformation $T_0$ and the articulation transformation $T_1$ are jointly optimized with a learning rate of 1e-2.
We initialize 10,000 Gaussian primitives for each OBB by uniformly sampling their positions within the canonical space $[-1,1]^3$. 

We use the following loss weights by default. The rendering loss weight is set to \( \lambda_{\text{image}} = 10.0 \). 
The part fitting loss consists of a coverage loss and a margin loss, with \( \lambda_{\text{coverage}} = 100.0 \) and \( \lambda_{\text{margin}} = 50.0 \), respectively. 
For the contact loss, \( \lambda_{\text{contact}} = 0.1 \). 
The extent regularizer is set with \( \lambda_{\text{ext}} = 0.01 \).

We optimize StructureGS using a loss schedule. The structure-aware losses and the extent regularizer are activated from the beginning to guide the initial construction of part structure. Starting from iteration 400, we enable the rendering loss to jointly optimize part appearance and geometry. Once the part structure stabilizes, the contact loss and extent regularizer are disabled at iteration 1,500, and subsequent refinement is performed using the rendering and part fitting losses.

All experiments were conducted on a workstation with an NVIDIA GeForce RTX 4090 GPU and an Intel Core i9-12900K CPU, using PyTorch 2.0.1~\cite{paszke2017automatic} with CUDA 11.8.  
We use the differentiable 3D Gaussian Splatting renderer provided by the \texttt{gsplat}~\cite{ye2025gsplat} library.  
The average training time per scene is approximately 13 minutes, varying with scene complexity.

\subsection{Densification strategy} 
We adopt the densification strategy of 3D Gaussian Splatting~\cite{kerbl20233d} during training. All Gaussians from all OBBs are rendered jointly, and Gaussians are split or duplicated based on the rendering error and opacity, following the original 3DGS framework~\cite{kerbl20233d}. To preserve Gaussian-to-OBB membership, new Gaussians are assigned to the same OBB as their parent Gaussian.

\subsection{OBB Initialization}
We initialize OBB parameters using coarse point clouds estimated from the two articulation states. Given point clouds $\mathcal{P}_0$ and $\mathcal{P}_1$ reconstructed by VGGT~\cite{wang2025vggt}, we compute a nearest-neighbor motion cue for each point $\mathbf{p}_i \in \mathcal{P}_0$:
\begin{equation}
d_i = \min_{\mathbf{q}_j \in \mathcal{P}_1} |\mathbf{p}_i - \mathbf{q}_j|_2 .
\end{equation}
We classify points with small motion cues as static and use the remaining points as moving candidates. Specifically, we threshold the distances ${d_i}$ by $\tau_{\text{motion}}$ to separate static and moving candidate points.

The moving candidate points are clustered using DBSCAN~\cite{ester1996density}. The resulting clusters are treated as moving-part hypotheses. If the number of clusters is larger than the expected number of moving parts, we keep the dominant clusters by size and remove small noisy clusters. We then fit an oriented bounding box to each part hypothesis, including the static region. These fitted boxes initialize the OBB transforms $T_0^k$. The articulation transforms $T_1^k$ are initialized to identity for all parts.

The initialization is intentionally coarse. Its role is to place the oriented boxes near plausible object parts, rather than to provide accurate segmentation. During optimization, the Gaussian primitives, OBB parameters, and articulation parameters are jointly refined by photometric and structure-aware losses.

\subsection{Joint type and parameter estimation} 
The motion of each part is obtained from its optimized part-wise OBB transformation \( \mathcal{T}_1^k \). For transformation \( \mathcal{T}_1^k \), we first determine the joint type and then recover the corresponding joint parameters.

To determine the joint type, we compute the rotation angle from the rotational component of \( \mathcal{T}_1^k \). A part is classified as \emph{prismatic} if the rotation angle is below a threshold of \( 5^\circ \) (i.e., \( \theta < 0.087 \) radians), and \emph{revolute} otherwise.

For prismatic joints, we estimate the motion axis by normalizing the translation vector. The joint displacement is measured as the \( \ell_2 \) norm of the translation.  
For revolute joints, we extract the rotation axis by converting the rotational component to axis-angle form. The pivot point corresponds to the fixed point of the rigid transformation, i.e., a point that remains unchanged under the transformation. We compute the pivot point as the minimum-norm solution to a linear system derived from \( (R - I)\,p = -t \), where \( R \) and \( t \) are the rotation and translation components of \( \mathcal{T}_1^k \).

\begin{figure}[t]
    \centering
    \includegraphics[width=\linewidth,trim=0cm 10cm 4cm 0cm, clip]{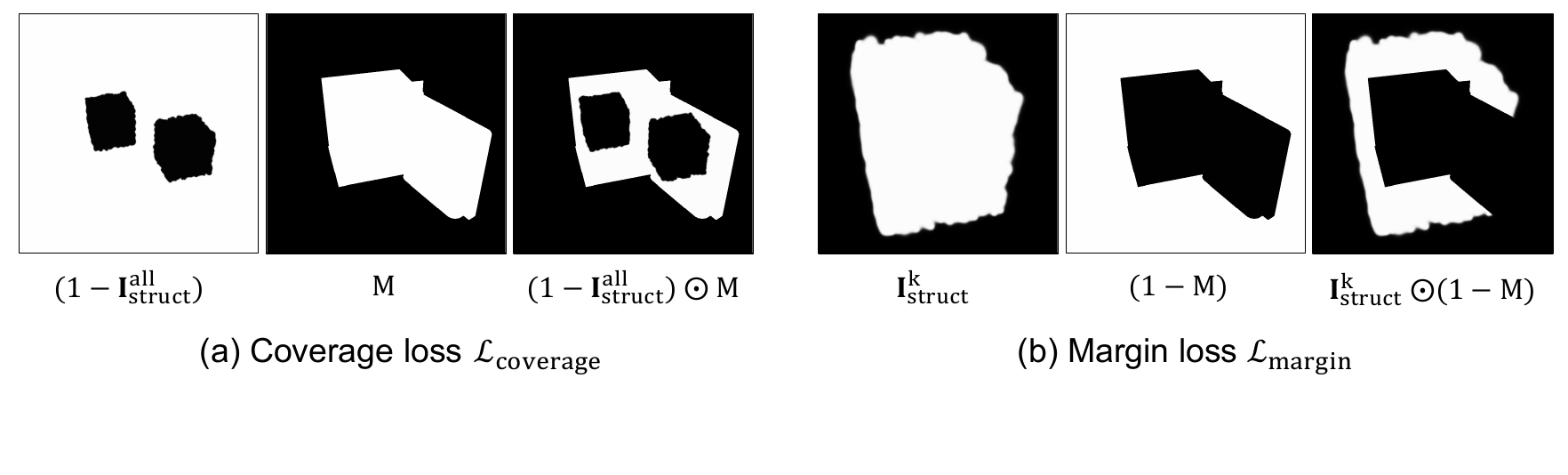} 
    \caption{Visualization of the error maps and intermediate terms of the part fitting loss. \(M\) denotes the binary ground-truth object mask, \(\mathbf{I}_{\text{struct}}^{k}\) the occupancy map of the \(k\)-th part OBB, and \(\mathbf{I}_{\text{struct}}^{\text{all}}\) their union. (a) and (b) illustrate two example cases where the coverage loss and the margin loss are non-zero, respectively. In (a), the error map is formed by multiplying the complement of the aggregated structure image \( (1-\mathbf{I}_{\text{struct}}^{\text{all}}) \) with the object mask \( M \), yielding \( (1-\mathbf{I}_{\text{struct}}^{\text{all}})\odot M \), which indicates object regions not covered by the union of OBBs.  In (b), the error map is formed by multiplying the per-part structure image \( \mathbf{I}_{\text{struct}}^k \) with the region outside the object mask \( (1-M) \), yielding \( \mathbf{I}_{\text{struct}}^k \odot (1-M) \), which indicates OBB regions outside the object. }
    \label{fig:loss_viz}
\end{figure}

\subsection{Error map for the part fitting loss} 
To provide an intuitive understanding of the part fitting loss  $\mathcal{L}_{\text{fit}}$ that consists of the coverage loss  $\mathcal{L}_{\text{coverage}}$ and the margin loss $\mathcal{L}_{\text{margin}}$, we visualize the error maps of the individual loss terms, together with the structure image and object mask used to define the loss terms.




The coverage loss becomes active when regions inside the object mask are not covered by the union of OBBs. In \Fig{loss_viz}a, we show the two intermediate terms, the complement of the all-part structure image ($1 - I_{\text{struct}}^{\text{all}}$)  and the object mask $M$, together with the resulting error map obtained by their element-wise product.
The error map highlights uncovered regions inside the object mask, encouraging the OBBs to fully cover the object.

The margin loss encourages each OBB to tightly fit the object. In \Fig{loss_viz}b, we visualize the part structure image \( I_{\text{struct}}^k \) and the complement of the object mask \( 1 - M \), together with the resulting error map obtained by their element-wise product. This error map captures the empty space inside the OBB that is not occupied by the object.



\begin{figure*}[t]
    \centering
        \includegraphics[width=\linewidth,trim=0cm 10cm 0cm 0cm, clip]{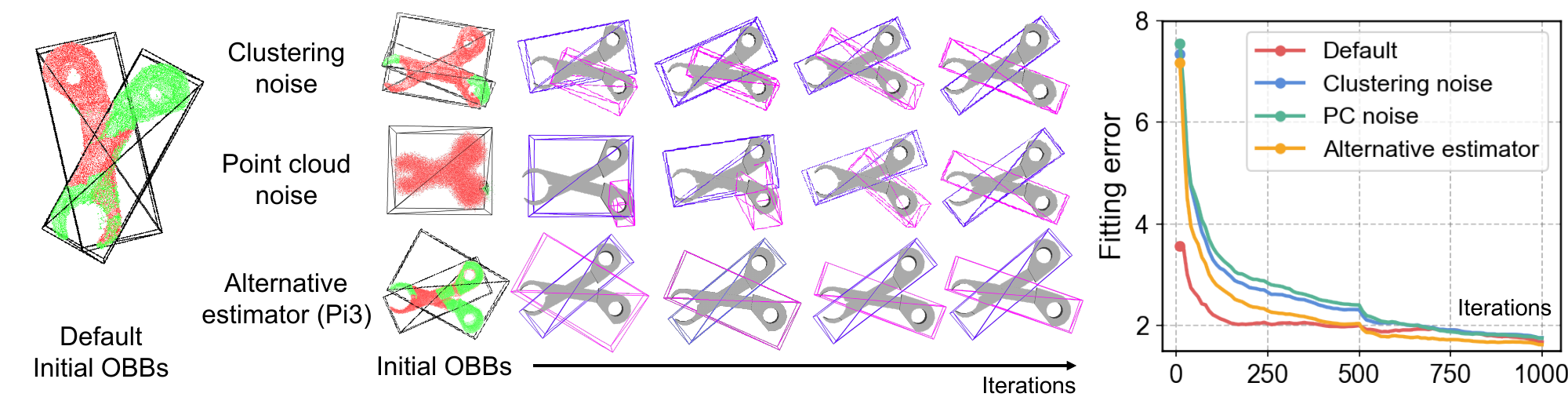}
    \caption{Left: Default initial OBBs obtained from motion-based point clustering, where point colors indicate clustered part points. Middle: OBB optimization results over iterations under clustering noise, point cloud noise, and an alternative point cloud estimator. Right: Fitting error curves over optimization iterations for the default and perturbed settings.  }
    \label{fig:init_impact} 
\end{figure*}
\section{Robustness to OBB initialization}
To initialize the OBB parameters, we first obtain point clouds of the target object in the initial and articulated states using VGGT~\cite{wang2025vggt}. We then cluster the points into static and dynamic parts based on the distances between corresponding points in the two states, and fit an OBB to each cluster. This initialization provides only coarse and noisy structural information, as it depends on imperfect point cloud estimation and point clustering. Nevertheless, our optimization remains robust to such noisy initialization and can still recover a reliable solution. We examine this robustness in the following experiment.

We evaluate robustness under three settings with 10 sparse views: clustering noise, point cloud noise, and an alternative point cloud estimator. Clustering noise is introduced by varying the threshold used to separate static and dynamic points across the two states, which reduces the accuracy of part segmentation and thus degrades the OBB initialization. Point cloud noise is introduced by adding Gaussian perturbations to the points estimated by VGGT~\cite{wang2025vggt} before performing part clustering for OBB initialization. Finally, we replace the default point estimator (VGGT) with a recent alternative method (Pi3~\cite{wang2026pi3permutationequivariantvisualgeometry}) to evaluate the robustness of our method to different point estimation methods. These perturbations produce OBB initializations of varying quality, ranging from reasonable part-wise boxes to severely imbalanced cases. For example, a single OBB covers most of the object while the remaining boxes capture only small fragments, as shown in the “Point cloud noise” of \Fig{init_impact}.

\Fig{init_impact} shows the initial OBBs and their evolution during optimization under these perturbation settings, together with the corresponding fitting error curves. Although the initial OBB quality varies substantially, the fitting error decreases during optimization and converges to similar values across all settings. Qualitatively, the OBBs progressively align with the object parts even when starting from different initializations.
These results demonstrate that our method can recover reliable part-aligned structures from inaccurate OBB initializations. Moreover, the comparable convergence obtained with Pi3~\cite{wang2026pi3permutationequivariantvisualgeometry} indicates that our method does not rely on a specific point cloud estimator.



\section{Additional Experimental Results}
\subsection{Per-scene quantitative comparison on the PARIS dataset} 
In the main paper, we report the average performance on the PARIS synthetic benchmark. Here, for completeness, we provide the per-scene quantitative results under dense-view and sparse-view settings, respectively (\Tbl{quantitative}).

Among the prior methods listed in Table 1 of the main paper, we compare our method only with those evaluated under comparable settings and metrics. Since our method takes multi-view RGB images of two states as input and does not assume known joint types, methods based on different input modalities or problem settings are excluded from the comparison. Specifically, A-SDF~\cite{mu2021sdf} is a data-driven learning-based method, REACTO~\cite{song2024reacto} requires monocular RGB video sequences, ArtGS~\cite{liu2025artgs} additionally relies on depth input, and SPLART~\cite{lin2025splart} assumes known joint types.

\subsection{Additional qualitative comparison on the PARIS dataset}
We also present additional qualitative comparisons on the PARIS dataset under both full-view and sparse-view settings. 
For each example, we reconstruct static and dynamic part meshes and visualize the estimated articulation axes.

\begin{figure}[t]
    \centering
    \includegraphics[width=\linewidth,trim=0cm 13cm 0cm 0cm, clip]{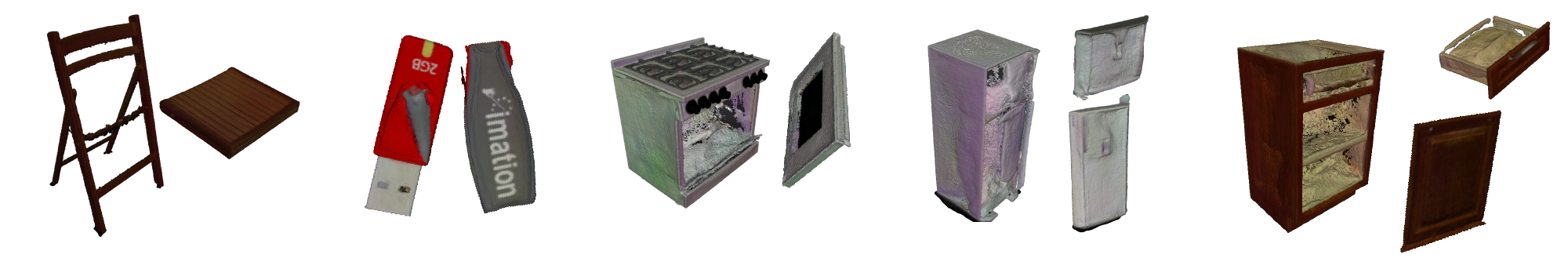}
    \caption{Gallery of part-wise reconstructed meshes across different object categories.}
    \label{fig:part_mesh}
    \vspace*{-0.2cm}
\end{figure}

As discussed in the main paper, existing methods generally exhibit degraded performance under sparse-view setting compared with the full-view setting. In contrast, our method remains robust in both sparse-view and full-view settings, achieving reliable part decomposition and articulation estimation.
We provide a gallery of reconstructed part meshes in \Fig{part_mesh} to further illustrate the geometric fidelity of the recovered results.

\begin{figure}[t]
    \centering
    \includegraphics[width=0.95\linewidth,trim=0cm 7cm 0cm 0cm, clip]{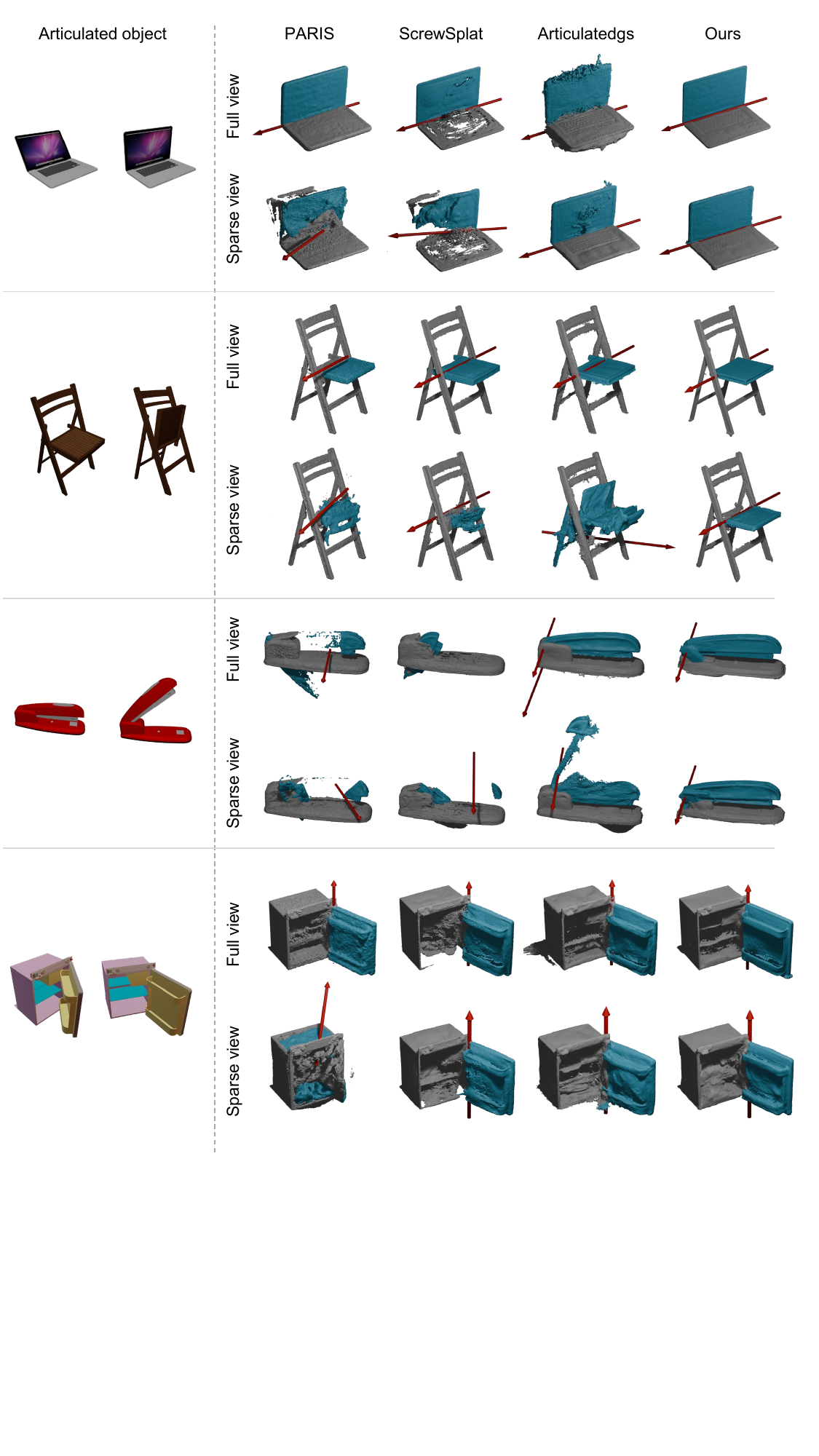} 
    \caption{Further qualitative comparison on the PARIS dataset~\cite{liu2023paris} under full-view and sparse-view settings.
The left column shows the articulated objects in two states. The remaining columns visualize reconstruction and articulation estimation results of different methods (PARIS~\cite{liu2023paris}, ScrewSplat~\cite{kim2025screwsplat}, ArticulatedGS~\cite{guo2025articulatedgs}, and ours). Dynamic parts are shown in blue and static parts in gray, while red arrows indicate the estimated articulation axes. }
    \label{fig:real_animation}
    \vspace*{-0.2cm}
\end{figure}

\subsection{Fitted OBBs and reconstructed 3DGS on the PARIS dataset} 
\begin{figure*}[t]
    \centering
    \includegraphics[width=0.95\linewidth,trim=0cm 13cm 0cm 0cm, clip]{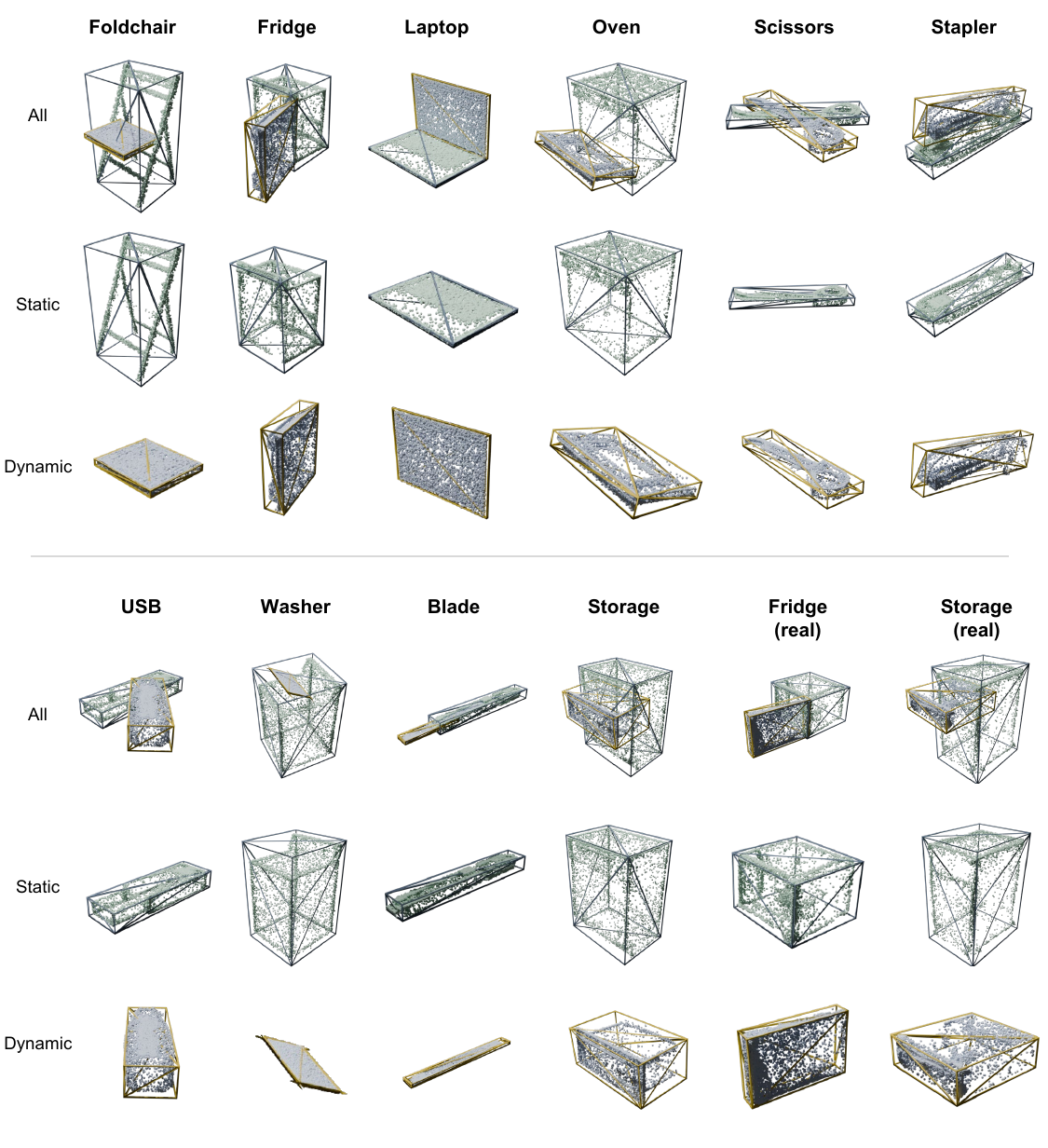} 
    \caption{ OBB and 3DGS fitting results on the PARIS dataset. Columns correspond to different scenes from the PARIS benchmark.  Rows show the full reconstruction (“All”) and the static and dynamic parts.
For each case, we visualize the fitted OBBs together with the reconstructed 3D Gaussians. }
    \label{fig:obb_pts} 
\end{figure*}

\Fig{obb_pts} shows additional results of our method on scenes from the PARIS benchmark. 
We visualize the fitted OBBs and reconstructed 3D Gaussians  for each part. 
Across all scenes, the jointly optimized Gaussians capture geometric details, while the OBBs represent the underlying part structure.
The results also demonstrate that our method generalizes well to parts with non-cuboidal geometries, as discussed in Section 5.4 of the main paper. For example, \textit{Scissors}, \textit{Stapler}, and \textit{Fridge} scenes include highly concave regions.

\subsection{Scalability to multiple moving parts on ArtGS-Multi dataset} 
To evaluate scalability beyond objects with a single moving part, we conduct additional experiments on ArtGS-Multi~\cite{liu2025artgs}, which contains articulated objects with 4 to 7 parts. Since ArtGS-Multi provides RGB-D observations, we use the provided depth only to construct the initial point clouds for OBB initialization, and do not use depth during optimization. We compare our method with ScrewSplat~\cite{kim2025screwsplat}, which supports reconstruction of articulated objects with multiple moving parts. As shown in \Tbl{artgs_multi}, our method successfully converges on multi-part objects and substantially improves both geometric and kinematic accuracy over the baseline, demonstrating that our part-level parameterization scales to more complex articulated structures.

\subsection{Animation of real-world reconstructions}
We provide animations for the real-world results in Sec.~5.3 of the main paper. For each example, we interpolate the motion of the dynamic parts using the estimated articulation parameters, transform the corresponding 3DGS set to intermediate states, and render the resulting images. These animations demonstrate that our method can robustly recover both part-wise geometry and articulation parameters.

\begin{table}[t]
\centering
\caption{ Quantitative results on the ArtGS-Multi dataset~\cite{liu2025artgs}. We compare with ScrewSplat~\cite{kim2025screwsplat}, a baseline that supports articulated reconstruction with multiple moving parts. Ang Err and Pos Err evaluate joint-axis accuracy, GD-r and GD-p measure motion errors for revolute and prismatic motions, respectively, and CD-s/CD-m/CD-w report Chamfer-L1 distances for the static base, movable parts, and whole object. Lower is better for all metrics.} 
\label{tbl:artgs_multi}
\small
\resizebox{0.7\columnwidth}{!}{
 \begin{tabular}{l cccc ccc}
        \toprule          
         & Ang Err & Pos Err & GD-r & GD-p & CD-s & CD-m & CD-w \\
        \midrule
        ScrewSplat & 27.4 & 0.13 & 54.7 & 0.01 & 58.2 & 684.3 & 66.2 \\
        Ours & 3.67 & 0.03 & 7.05 & 0.00 & 2.53 & 2.70 & 2.64 \\
        \bottomrule
      \end{tabular}
}
\end{table}

\begin{table}[t] 
\centering
\caption{  
Quantitative per-scene results on the PARIS dataset under full-view and sparse-view settings with 100 and 10 input views, respectively. Geometry is measured by Chamfer Distance on the whole object (CD-w), static part (CD-s), and movable part (CD-m), with all values scaled by $10^3$. Motion is evaluated by joint axis angular error, position error for revolute joints, and motion magnitude via geodesic and translation errors for revolute and prismatic motions, respectively. ``F'' denotes failure to predict a valid motion type, and ``--'' indicates non-applicable metrics.  
Per-scene motion metrics can be interpreted in the context of their scales: in most scenes under both full-view and sparse-view settings, angular errors are within $1^\circ$, position errors measured in meters are below 0.01\,m, and geodesic errors are near $1^\circ$. These scene-wise variations are therefore numerically marginal. More importantly, our method achieves consistently low geometric error, whereas other methods may obtain competitive motion estimates while degrading geometric accuracy.
}
\vspace*{-0.5cm}
\label{tbl:quantitative}
\begin{subtable}{\linewidth}
\caption{
Full-view setting (100 views)
}
\vspace*{-0.2cm}
\footnotesize               
\setlength{\tabcolsep}{5pt} 
\renewcommand{\arraystretch}{1} 
\resizebox{\textwidth}{!}{%
\begin{tabular}{lllcccccccccccc}
\toprule
&&& \multicolumn{10}{c}{Synthetic} & \multicolumn{2}{c}{Real} \\
\cmidrule(lr){4-13}\cmidrule(lr){14-15}
Group & Metric & Method & Foldchair & Fridge & Laptop & Oven & Scissor & Stapler & USB & Washer & Blade & Storage & Fridge & Storage  \\
\midrule
\multirow{12}{*}{Motion}
  & \multirow{4}{*}{Ang Err}    & PARIS         &   0.837       & 0.128 & 0.867 & 0.204 & 2.451 & 1.059 & 0.179 & 2.323 & 48.118 & 0.179 & 2.595 & 19.53  \\
  &                             & ScrewSplat    &  0.236        & 0.052 & 70.178 & 0.584 & \textbf{0.020} & 6.522 & 0.119 &\textbf{ 0.088} & 0.040 & 0.034 & 2.477 & \textbf{4.244}  \\ 
  &                             & ArticulatedGS &  0.049        & \textbf{0.028} & 0.152 & 0.06 & 0.066 & \textbf{0.074} & 0.162 & 0.279 & \textbf{0.020} & \textbf{0.021} & 3.13 & 85.21 \\
  &                             & Ours          &\textbf{0.019} & 0.055& \textbf{0.083} & \textbf{0.019} & 0.092 & 0.128 & \textbf{0.059} & 0.096 & 0.292 &  0.034  & \textbf{0.988} &  13.01 \\
\cmidrule(lr){2-13}\cmidrule(lr){14-15}  
  & \multirow{4}{*}{Pos Err}    & PARIS         &   0.218       & 0.005 & 0.009 & 0.001 & 0.361 & 1.020 & 0.832 & 0.636 & - & - &  0.016 & -  \\
  &                             & ScrewSplat    &  0.364        & 0.005 & 0.307 & 0.035 & 0.001 & 0.234 & 0.161 & 0.002 & - & - & 0.599 & -  \\
  &                             & ArticulatedGS &   0.001       & 0.002 & 0.061 & 0.003 & 0.002 & \textbf{0.001} & 0.072 & \textbf{0.000} & - & - & 0.043 & - \\  
  &                             & Ours          &\textbf{0.000} & \textbf{0.000} & \textbf{0.000} & \textbf{0.000} & \textbf{0.000} & 0.003 & \textbf{0.001} & \textbf{0.000} & - & - & \textbf{0.032} & -  \\
\cmidrule(lr){2-13}\cmidrule(lr){14-15}  

  & \multirow{4}{*}{Geo Dist}   & PARIS         &  177.068      & 0.256 & 1.002 & 0.601 & 138.20 & 129.11 & 179.66 & 66.23 & 1.523 & 0.602 & 5.893 & 0.541  \\
  &                             & ScrewSplat    &  14.602       & 0.555 & 47.378 & 8.930 & 79.942 & 20.378 & 85.726 & 0.069 & 0.400 &\textbf{0.000 }& 91.809 &  0.499  \\ 
  &                             & ArticulatedGS &  0.063        & 0.383 & 4.970 & 0.163 & \textbf{0.056} & \textbf{0.063} & 0.173 & 0.220 & 0.400 & \textbf{0.000} & 6.195  & F \\ 
  &                             & Ours          &\textbf{0.027} & \textbf{0.068} & \textbf{0.055} &\textbf{0.027} & 0.062 & 0.163 & \textbf{0.068} & \textbf{0.055} &\textbf{ 0.001} & \textbf{0.000} &\textbf{ 0.017} &  \textbf{0.024}  \\
\midrule
\multirow{12}{*}{Geometry}
  & \multirow{4}{*}{CD-s}       & PARIS         &   11.676      & 3.064 & \textbf{0.202} & 9.485 & 2.636 & \textbf{1.718} & 2.189 & 16.291 & 1.806 & 6.206 &  41.897  & 67.321 \\
  &                             & ScrewSplat    &  13.078       & 46.14 & 48.765 & 45.694 & 131.813 & 32.154 & 31.071 & 31.389 & 0.584 & 8.816 &  138.55  & 8.29   \\
  &                             & ArticulatedGS &   3.713       & 1.674 & 1.257 & 2.185 & 0.342 & 1.735 & 1.927 & 5.364 & \textbf{0.304} & 2.687 & 33.02  & 312.78  \\ 
  &                             & Ours          &\textbf{0.358} & \textbf{0.880} & 0.374 &\textbf{1.503} & \textbf{0.279} & 5.360 & \textbf{0.860} & \textbf{4.083} & 0.311 & \textbf{1.576} &  \textbf{1.883}  &\textbf{ 3.06 }\\
\cmidrule(lr){2-13}\cmidrule(lr){14-15}  

  & \multirow{4}{*}{CD-m}       & PARIS         &   11.57      & 1.957 & 0.236 & 69.32 & 23.75 & 121.71 & 7.513 & 261.10 & F & 91.10 &  119.60  & 293.69 \\
  &                             & ScrewSplat    &  34.295       & 1.433 & 17.723 & 126.36 & 0.257 & 61.70 & 2.160 & 43.82 & 2.636 & 31.617 &  12.39  & 50.76  \\ 
  &                             & ArticulatedGS &\textbf{0.520} & 0.699 & 6.05 & 0.818 & 0.407 & 1.283 & 1.061 & \textbf{1.833} & 1.526 & 1.664 &   71.61  & 1122.6  \\ 
  &                             & Ours          & 1.083         & \textbf{0.562} & \textbf{0.173} & \textbf{0.309} &\textbf{0.249} & \textbf{1.123} & \textbf{0.663} & 7.639 & \textbf{1.083} & \textbf{1.113}   & \textbf{0.841}  &\textbf{ 15.55}  \\
\cmidrule(lr){2-13}\cmidrule(lr){14-15}  

  & \multirow{4}{*}{CD-w}       & PARIS         &   1.238       & 2.398 & \textbf{0.228} & 5.940 & 1.427 & 19.03 & 1.341 & 20.07 & 0.547 & 6.722 &  17.59  & 47.11 \\
  &                             & ScrewSplat    &  1.986        & 4.124 & 6.004 & 5.444 & 45.28 & 25.66 & 15.45 & 22.204 & 0.425 & 12.78 &  20.04  & 8.77 \\ 
  &                             & ArticulatedGS &  0.512        & 1.395 & 5.010 & 2.024 & 0.331 & 1.508 & 1.265 & 4.918 & 0.231 & 1.890 & 23.09  &  231.41 \\            
  &                             & Ours          &\textbf{0.227} & \textbf{0.847} & 0.281 & \textbf{1.368} & \textbf{0.248} & \textbf{0.881} & \textbf{0.745} & \textbf{2.985} & \textbf{0.175} & \textbf{1.441}  &\textbf{ 1.242}  & \textbf{3.226} \\
\bottomrule
\end{tabular}} 
\end{subtable}
\begin{subtable}{\linewidth}
\centering
\vspace*{0.5cm}
\caption{
Sparse-view setting (10 views)
}

\vspace*{-0.2cm}
\footnotesize               
\setlength{\tabcolsep}{5pt} 
\renewcommand{\arraystretch}{1} 
\resizebox{\textwidth}{!}{%
\begin{tabular}{lllcccccccccccc}
\toprule
&&& \multicolumn{10}{c}{Synthetic} & \multicolumn{2}{c}{Real} \\
\cmidrule(lr){4-13}\cmidrule(lr){14-15}
Group & Metric & Method & Foldchair & Fridge & Laptop & Oven & Scissor & Stapler & USB & Washer & Blade & Storage & Fridge & Storage  \\
\midrule
\multirow{12}{*}{Motion}
  & \multirow{4}{*}{Ang Err} & PARIS            &  22.03  & 27.92 & 27.95 & 75.54 & 4.15 & 30.15 & 1.08 & 18.19 & 66.97 & 54.39 &  34.06 & 46.33 \\
  &                         & ScrewSplat        &   \textbf{0.138} &  \textbf{0.034} & 11.167 & 6.757 & 0.059 & 8.085 & 1.324 & 87.14 & 0.335 &  \textbf{0.32} & 2.192 & 89.08  \\ 
  &                         & ArticulatedGS     & 51.74  &  0.101 &\textbf{ 0.138} & 1.922 & \textbf{ 0.04} & 3.448 & 0.459 & 9.357 & \textbf{0.034} & 28.854 & 4.575 & 35.55   \\ 
  &                         & Ours              & 0.935 & 1.061 & 0.280 &\textbf{ 0.506} & 0.117 &  \textbf{0.911} &  \textbf{0.099} &  \textbf{0.466} & 0.852 & 1.786 &  \textbf{1.25} &   \textbf{12.25}  \\
\cmidrule(lr){2-13}\cmidrule(lr){14-15}  
  & \multirow{4}{*}{Pos Err} & PARIS            & 0.137   &  0.286 & 0.131 & 0.050 & 0.180 & 0.403 & 0.004 & 0.110 & - & - & 0.211 & -  \\
  &                         & ScrewSplat &      \textbf{ 0.001}    &\textbf{  0.001}  & 0.007 & 0.056 & 0.003 & 0.81 & 0.391 & F & - & - & 0.066 & -   \\
  &                         & ArticulatedGS     & 0.592 & 0.014 & 0.003 & 0.378 & \textbf{0.000} & 0.428 & \textbf{ 0.000}  & 0.028 & - & - & \textbf{ 0.002 } & -  \\
  &                         & Ours              & \textbf{ 0.001 } & 0.024 &\textbf{  0.001}  &\textbf{  0.011}  &\textbf{  0.000 } &\textbf{  0.004}  & 0.001 & \textbf{ 0.014}  & - & - & 0.034 & - \\
\cmidrule(lr){2-13}\cmidrule(lr){14-15}  

  & \multirow{4}{*}{Geo Dist} & PARIS           &  102.81 & 74.62  & 28.7 &  92.1& 82.91 & 104.23 & 0.97 & 88.38 & 0.090 & 0.19 & 24.2 & 0.18 \\
  &                          & ScrewSplat &  \textbf{  0.206  } & 119.93 & 80.711 & 5.081 & 0.069 & 52.30 & 21.43 & F & 0.401 &  F & 1.632 & F  \\ 
  &                          & ArticulatedGS    & 79.66 & 4.271 & 0.384 & 11.96 & \textbf{ 0.040 } & 69.50 & 0.388 & 17.03 & 0.400  & 0.141 & 4.968 & 0.642   \\ 
  &                          & Ours             & 1.207 &\textbf{  4.121 } &\textbf{  0.267 } &\textbf{3.531} & 0.088 & \textbf{ 0.626  } & \textbf{ 0.088  }& \textbf{ 1.046  } & \textbf{ 0.080  }& \textbf{ 0.004  }&\textbf{  0.927  }& \textbf{ 0.050   }  \\
\midrule
\multirow{12}{*}{Geometry}
  & \multirow{4}{*}{CD-s} & PARIS               &  309.16 & 180.69 &92.14  & 34.21 &907.8  & 3.16  & 2.46 &16.34  & 1.12 &27.43 &  63.28 & 55.08   \\
  &                      & ScrewSplat &            \textbf{ 1.72}    & 6.95 & 132.85 & 14.39 & 33.75 & 99.51 & 49.46 & 43.01 &1.2  & 38.32  &   129.28 & 24.58   \\
  &                      & ArticulatedGS        & 2.856 & \textbf{0.993} & 1.121 & 2.522 & 0.640 & \textbf{1.491} & 1.67 &  5.723 & \textbf{0.215} & 3.218 &  38.31 &  242.27    \\
  &                      & Ours                 &  3.483  & 1.256 & \textbf{0.297} &\textbf{ 1.749} &\textbf{ 0.326 }& 3.708 & \textbf{0.749 }&\textbf{ 5.046 }& 0.348 &\textbf{ 2.606} &\textbf{2.211 }& \textbf{3.168}\\
\cmidrule(lr){2-13}\cmidrule(lr){14-15}  

  & \multirow{4}{*}{CD-m} & PARIS                & 98.13   & 564.73 & 130.76 & 236.95 & 141.4 & 94.94 & \textbf{2.47} & 62.73 &  F & 158.55 &  173.3 & 875.93 \\
  &                      & ScrewSplat           &  88.27  & 117.64 & 3.51 & 436.79 & 37.14 & 70.42 & 123.1 & 123.1 & 105.52 & 229.52 & 34.39 & 36.55 \\ 
  &                      & ArticulatedGS        &   104.30 & 1.434 & 0.693 & 22.69  & 0.568 & 445.56 & 3.810 & \textbf{ 0.212} & 3.423& 44.303 & 66.40 & 931.62  \\ 
  &                      & Ours                 &  \textbf{5.244 } &\textbf{ 0.890} & \textbf{0.151} & \textbf{0.508} &\textbf{ 0.288 }& \textbf{1.071 }& 3.263 & 3.909 &\textbf{ 0.967 }& \textbf{3.522} &  \textbf{1.086} & \textbf{22.127}\\
\cmidrule(lr){2-13}\cmidrule(lr){14-15}  

  & \multirow{4}{*}{CD-w} & PARIS               & 175.73 &  198.72 & 82.82  & 32.47  &  312.9 & 13.45 & 2.11 & 18.37& 0.56  & 9.2  &  32.24 & 65.58 \\
  &                      & ScrewSplat           &   3.99 & 63.04 & 35.32 & 26.35 & 21.38 & 43.33 & 16.14 & 16.14 & 55.02 & 30.21  &  15.58 & 12.96  \\ 
  &                      & ArticulatedGS        & 2.837   &\textbf{ 1.024} & 0.617 & 2.577 & 0.583 & 181.03 & 1.156 &  5.278 & \textbf{0.195} & 3.158 &  25.14 &202.28  \\ 
  &                      & Ours                 &  \textbf{1.560} & 1.166 & \textbf{0.251} &\textbf{ 1.596 }& \textbf{0.302} &\textbf{ 0.714} & \textbf{1.011} & \textbf{3.935} & 0.204 &\textbf{ 2.462} & \textbf{ 1.509} &  \textbf{3.431 } \\
\bottomrule
\end{tabular}} 
\end{subtable}
\end{table}

\begin{figure}[t]
    \centering
    \includegraphics[width=\linewidth,trim=0cm 0cm 7.5cm 0cm, clip]{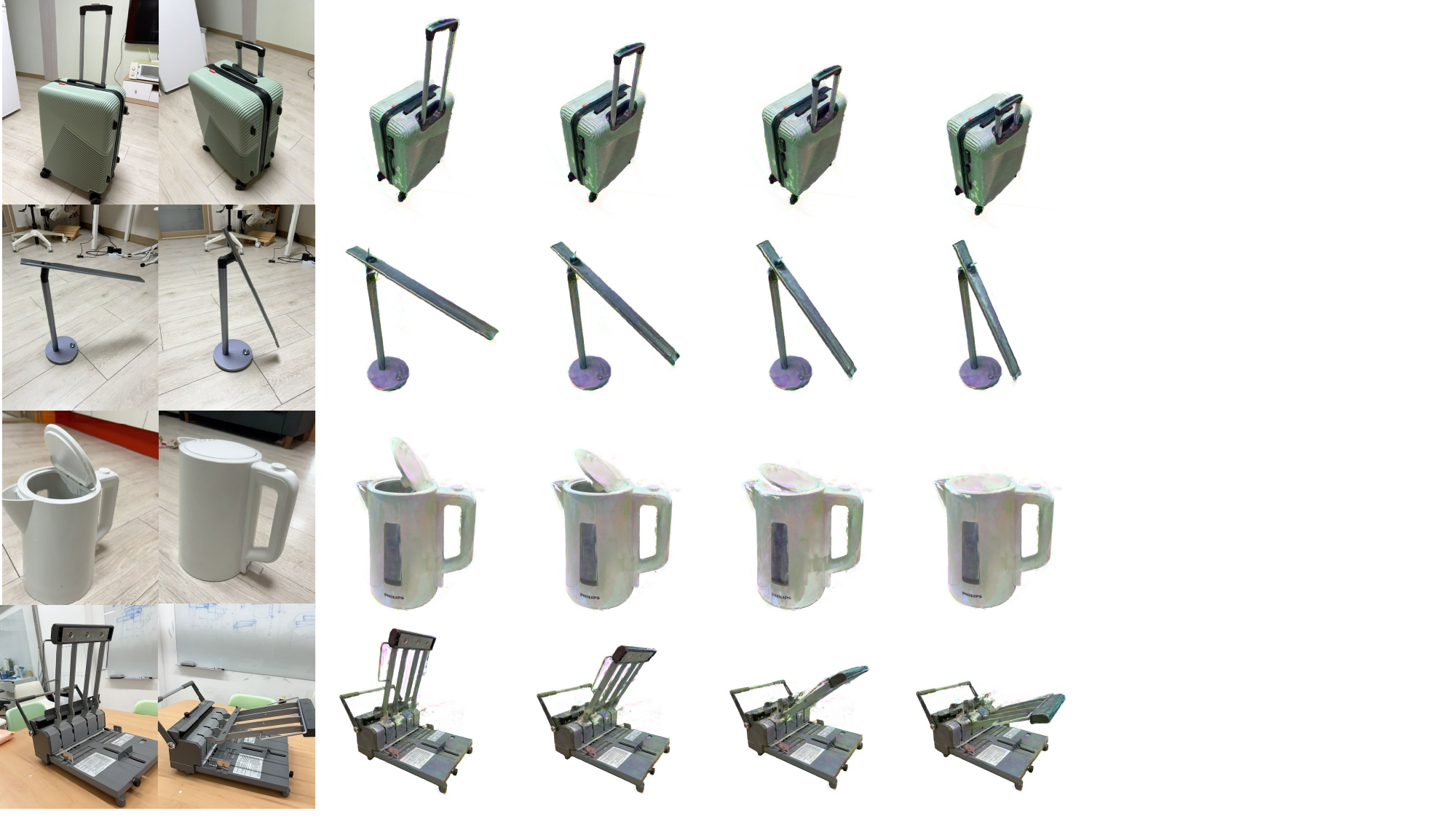} 
    \caption{Animation results on real-world objects.
For each example, the left two columns show the input image examples at the initial and articulated states, respectively. The remaining columns show rendered frames from the resulting animation, where the reconstructed part-level Gaussians are transformed according to the estimated articulation parameters.  }
    \label{fig:real_animation}
    \vspace*{-0.2cm}
\end{figure}
 
